\documentclass[11pt, a4paper, final]{article}
\usepackage{amsmath}
\usepackage{amsfonts}
\usepackage{amssymb}
\usepackage{amsthm}
\usepackage{mathrsfs}
\usepackage{mathtools}
\usepackage{dsfont}
\usepackage{graphicx}
\usepackage{color}
\usepackage[round]{natbib}

\usepackage[titletoc]{appendix}
\usepackage{bm}
\usepackage{algpseudocode}
\usepackage{algpascal}
\usepackage{afterpage}
\usepackage{enumitem}
\usepackage{multirow}
\usepackage{booktabs}

\usepackage{graphicx}
\usepackage{graphics}
\usepackage{pict2e}
\usepackage{epic}

\usepackage{epstopdf} 
\usepackage[ruled,vlined]{algorithm2e}
\usepackage{float}
\usepackage{tikz}
\usepackage{subfigure}
\usetikzlibrary{positioning, arrows, calc}

\usepackage[backref=page]{hyperref}
\hypersetup{
    colorlinks=true,
    citecolor=blue,
    linkcolor=red,
    filecolor=blue,      
    urlcolor=blue,
    pdfpagemode=FullScreen,
    }
\newlength{\vslength}
\newtheorem{theorem}{Theorem}[section]

\newtheorem{lemma}{Lemma}[section]

\newtheorem{proposition}{Proposition}

\newtheorem{corollary}{Corollary}[section]

\theoremstyle{remark}

\newtheorem{example}{\bf Example}[section]

\newcommand{\bbR}{{\mathbb R}}

\newcommand{\ie}{{\it i.e.}}

\newcommand{\bc}{\begin{center}}
\newcommand{\ec}{\end{center}}
\newcommand{\be}{\begin{equation}}
\newcommand{\ee}{\end{equation}}
\newcommand{\ba}{\begin{array}}
\newcommand{\ea}{\end{array}}
\newcommand{\bean}{\setlength\arraycolsep{1pt}\begin{eqnarray*}}
\newcommand{\eean}{\end{eqnarray*}}
\newcommand{\bea}{\setlength\arraycolsep{1pt}\begin{eqnarray}}
\newcommand{\eea}{\end{eqnarray}}
\newcommand{\ben}{\begin{enumerate}}
\newcommand{\een}{\end{enumerate}}
\newcommand{\bed}{\begin{itemize}}
\newcommand{\eed}{\end{itemize}}

\def\log{\text{log }}

\def\boxit#1{\vbox{\hrule\hbox{\vrule\kern6pt
 \vbox{\kern6pt#1\kern6pt}\kern6pt\vrule}\hrule}}

\def\bse{\begin{eqnarray*}}
\def\ese{\end{eqnarray*}}
\def\be{\begin{eqnarray}}
\def\ee{\end{eqnarray}}
\def\bq{\begin{equation}}
\def\eq{\end{equation}}
\def\bse{\begin{eqnarray*}}
\def\ese{\end{eqnarray*}}

\def\boxit#1{\vbox{\hrule\hbox{\vrule\kern6pt
          \vbox{\kern6pt#1\kern6pt}\kern6pt\vrule}\hrule}}

\def\log{\text{log }}

\def\bse{\begin{eqnarray*}}
\def\ese{\end{eqnarray*}}
\def\be{\begin{eqnarray}}
\def\ee{\end{eqnarray}}
\def\bq{\begin{equation}}
\def\eq{\end{equation}}
\def\bse{\begin{eqnarray*}}
\def\ese{\end{eqnarray*}}

\newcommand{\ua}       {\boldsymbol{a}}

\newcommand{\ub}       {\boldsymbol{b}}
\newcommand{\uC}       {\boldsymbol{C}}
\newcommand{\uc}       {\boldsymbol{c}}

\newcommand{\ug}       {\mbox{\boldmath$g$}}

\newcommand{\uR}       {\boldsymbol{R}}

\newcommand{\uS}       {\boldsymbol{S}}
\newcommand{\us}       {\boldsymbol{s}}

\newcommand{\uu}       {\boldsymbol{u}}

\newcommand{\ux}       {\boldsymbol{x}}

\newcommand{\uy}       {\boldsymbol{y}}
\newcommand{\uZ}       {\boldsymbol{Z}}
\newcommand{\uz}       {\boldsymbol{z}}

\newcommand{\ualpha}            {\mbox{\boldmath$\alpha$}}
\newcommand{\ubeta}             {\mbox{\boldmath$\beta$}}

\newcommand{\uepsilon}          {\mbox{\boldmath$\epsilon$}}

\newcommand{\ueta}              {\mbox{\boldmath$\eta$}}
\newcommand{\utheta}            {\mbox{\boldmath$\theta$}}

\newcommand{\uiota}             {\mbox{\boldmath$\uiota$}}

\newcommand{\uxi}               {\mbox{\boldmath$\xi$}}

\newcommand{\upi}               {\mbox{\boldmath$\pi$}}

\newcommand{\uphi}              {\mbox{\boldmath$\phi$}}

\newcommand{\upsi}              {\mbox{\boldmath$\psi$}}

\numberwithin{equation}{section}
\usepackage[margin=1in]{geometry}

\usepackage{comment}

\graphicspath{{Figures/}}

\makeatletter

\begin{document}
\thispagestyle{empty}
\title{
    \vspace*{-9mm}
    A monotone single index model for spatially referenced \\
    \vspace*{-3mm}multistate current status data
}
\author{\normalsize Snigdha Das$^{1}$, Minwoo Chae$^{2}$, Debdeep Pati$^{3}$, Dipankar Bandyopadhyay$^{4}$ \\[3mm]
    {\small\it {}$^{1}$Department of Statistics, Texas A$\&$M University} \\
    {\small\it {}$^{2}$Department of Industrial and Management Engineering,}
    \vspace{-0.09in} \\
    {\small\it  Pohang University of Science $\&$ Technology}\\
    {\small\it {}$^{3}$Department of Statistics, University of Wisconsin--Madison}\\
    {\small\it {}$^{4}$Department of Biostatistics,
Virginia Commonwealth University}
}

\date{}
\maketitle

\vspace*{-3mm}
\begin{abstract}
Assessment of multistate disease progression is commonplace in biomedical research, such as, in periodontal disease (PD). However, the presence of multistate current status endpoints, where only a single snapshot of each subject’s progression through disease states is available at a random inspection time after a known starting state, complicates the inferential framework. In addition, these endpoints can be clustered, and spatially associated, where a group of proximally located teeth (within subjects) may experience similar PD status, compared to those distally located. Motivated by a clinical study recording PD progression, we propose a Bayesian semiparametric accelerated failure time model with an inverse-Wishart proposal for accommodating (spatial) random effects, and flexible errors that follow a Dirichlet process mixture of Gaussians. For clinical interpretability, the systematic component of the event times is modeled using a monotone single index model, with the (unknown) link function estimated via a novel integrated basis expansion and basis coefficients endowed with constrained Gaussian process priors. In addition to establishing parameter identifiability, we present scalable computing via a combination of elliptical slice sampling, fast circulant embedding techniques, and smoothing of hard constraints, leading to straightforward estimation of parameters, and state occupation and transition probabilities. Using synthetic data, we study the finite sample properties of our Bayesian estimates, and their performance under model misspecification. We also illustrate our method via application to the real clinical PD dataset. 

\medskip
    
\noindent {\bf Keywords:} Accelerated failure time model; circulant embedding; clustered data; Dirichlet process; state occupation probability; transition probability. 
\end{abstract}


\section{Introduction}
\label{s:intro}

Multistate models have gained substantial momentum in biomedical research for event history analysis \citep{hougaard1999multi}, where a subject (or a study unit) can transition reversibly or irreversibly through a succession of intermediate states (reflective of the disease status under consideration) via a longitudinally observed process, before reaching an absorbing state (say, death). Similar to the usual survival data framework, the time-to-events are incomplete (or right-censored), and one observes the transition times and the corresponding states up to a random censoring time for each study unit. 

The current literature is inundated with frequentist \citep{wan2016continuous, Gu2023} and Bayesian \citep{Koslovsky2018} approaches to multistate models for interval-censored data \citep{vandenhout2016multi-state}, where subjects are followed longitudinally at multiple time points and the exact transition times are unobserved, but is known to occur within intervals defined by successive observations. However, it is relatively sparse for multistate models under Type-1 interval-censoring, or `current status' endpoints \citep{lan2017nonparametric, anyaso2023pseudo, anyasosamuel2024nonparametric}, 
where each subject is observed at a single inspection time after a known starting state, so that the individual’s state is known at two time points: the initial (starting) time and the inspection time, with no further follow-up. Hence, current status data is a special case of interval-censored data, and arises naturally in clinical trials, epidemiological studies, and surveys. Unlike longitudinal data, which provides detailed individual-level information on transitions between states over time, current status data offers limited insight on transitions due to its cross-sectional nature. This limitation necessitates specialized modeling approaches, as discussed in Chapter 8.2 of \cite{vandenhout2016multi-state}.

The motivation for this work comes from a cross-sectional study assessing periodontal disease (PD) status and progression among \textbf{G}ullah-speaking \textbf{A}frican-\textbf{A}merican Type-2 \textbf{D}iabetics \citep[henceforth, GAAD;][]{fernandes2009periodontal}. Here, the PD progression can be characterized as a multistate model, where the PD status pertaining to each tooth of a patient is classified into multiple states, according to PD severity, and are current status in nature. The modeling framework is further exacerbated by the presence of event times that are (a) correlated (current status tooth event times are clustered within subjects), and (b) with possible spatial association, since PD status of proximally-located teeth are hypothesized to be similar, and varying with regards to distally-located teeth \citep{reich2010latent}. Furthermore, (c) existing literature on multistate regression lacks the development of an \textit{interpretable} index summarizing the overall risk status of a subject. Specifically, \cite{lan2017nonparametric} and \cite{anyaso2023pseudo} developed univariate and multivariate regression approaches, respectively, accounting for informative cluster sizes and present risk estimates separately for each covariate, while \cite{anyasosamuel2024nonparametric} focus on the conditional estimation of future state occupation probabilities given prior occupation of specific states.
The objective of this paper is to develop a clinically interpretable index within a multistate framework  using a novel Bayesian nonparametric model with spatial random effects that account for the intraoral structure of teeth. The index combines the effect of all the risk factors into a single-number summary that informs and predicts the probabilities of occupying and transitioning through the PD states over time. However, note that the generic scientific question addressed in this work extends beyond one specific disease area (PD). Similar data structure and associated challenges, such as multistate models with current status endpoints, correlated pattern, and spatial association, can be observed in other biomedical domains, such as in cancer screening (breast/colorectal), infectious disease progression (COVID-19, tuberculosis), and organ transplantation (kidney graft survival).

In the absence of a unified framework addressing challenges (a–c) for developing an interpretable index, we adopt a semiparametric accelerated failure time (AFT) regression approach \citep{henderson2020individualized} for the time to the absorbing disease state, motivated by our naive preliminary analysis (presented in Section \ref{s:data_prelim} of the Supplement).
The latent survival times are linked to the observed state occupations by relative increments of time to each state that are Dirichlet distributed. Our hierarchical Bayes framework is the first of its kind, offering several innovations.  \textit{First}, we move away from the traditional Cox proportional hazards regression framework, where violations of the proportional hazards assumption may lead to imprecise inference, and further relax the (restrictive) parametric assumptions of the AFT errors by assigning a Dirichlet process mixture prior \citep{ sethuraman1994constructive} of Gaussians. \textit{Second}, the systematic component of the survival time is modeled through a flexible and clinically interpretable \textit{monotone} single index model \citep[SIM,][]{balabdaoui2019msim}, with the unknown link function estimated via a novel integrated basis expansion \citep{maatouk2017gaussian} and constrained Gaussian process priors on the basis coefficients. Posterior sampling is performed using the algorithm of \cite{Ray2020EfficientBS}, which uses a combination of the efficient elliptical slice sampling \citep{murray2010elliptical}, fast circulant embedding techniques \citep{wood1994simulation} and smooth relaxation of hard constraints. \textit{Third}, the intraoral spatial association among teeth for each subject is accounted for via multivariate Gaussian random effects, with an inverse-Wishart prior on its covariance matrix that is centered around a conditionally autoregressive model encouraging association between neighboring teeth. \textit{Finally},  our model allows for uncertainty quantification of disease state occupation probabilities (SOPs) and transition probabilities (TPs) over time, with individual trajectories described by the single index and thus interpretable in terms of the risk factors for PD.

The rest of the paper proceeds as follows. Section \ref{s:method} presents the statistical model, and develops the Bayesian inferential framework. 
While Section \ref{s:data_analysis} illustrates the model via application to the motivating GAAD data, Section \ref{s:sims} conducts simulation studies to assess finite sample properties of our model estimates, and the robustness of our proposed model under misspecification. Finally, Section \ref{s:conc} concludes, alluding to some future work.

\section{Statistical methodology}
\label{s:method}

\subsection{Model and likelihood}
\label{s:model}

To proceed with model building, we begin with setting up the necessary notations. Let $i$, $j$ and $k$ denote the subjects, teeth for each subject and the disease state occupied by each tooth, respectively, with  $i = 1,2,\ldots, n$, $j = 1,2,\ldots, m$, $k = 0,1,\ldots, K$. For tooth $j$ in subject $i$, let $S_{ij}(t)$ denote the disease state occupied at time $t \geq 0$, $C_{ij}$ denote the random time of inspection, and $\ux_{ij}$ denote the $p$-dimensional vector of covariates. We consider a progressive multistate model having $K+1$ states, where the transition of PD happens over time only in one direction, $0 \rightarrow 1 \rightarrow \ldots \rightarrow K$. The last state, $K$, is an absorbing state, meaning that once this state is reached, no further transitions occur. Let $\mathcal{K} = \{0,1,\ldots, K\}$ denote the finite state space of our multistate process having the directed transition structure. The observed CS data is represented as $\{ \ux_{ij}, C_{ij}, S_{ij}(C_{ij})\}$, where the inspection time $C_{ij}$ (which is the same as the censoring time for CS data) is independent of the continuous-time multistate process $\{S_{ij} (t) : t \geq 0\}$ given the covariate $\ux_{ij}$, under the assumption of independent censoring.

At the initial time $t = 0$, we assume that each tooth of every subject begins in state $0$. Let $T_{ij}^{\,(k)}$ represent the time elapsed (from the initial time) until the tooth reaches disease state $k$, where $T_{ij}^{\,(0)} = 0$. Defining $T_{ij}^{\,(K+1)} = \infty$, the disease state for tooth $j$ of subject $i$ at a given time $t \geq 0$ is given by
\begin{equation}
\label{eq:states}
S_{ij}(t) = k, \quad \textrm{if}\quad  T_{ij}^{\,(k)} \leq t < T_{ij}^{\,(k+1)}, \quad k = 0,1,\ldots, K.
\end{equation}

\begin{figure}
\begin{center}
    \subfigure[]{\scalebox{1.9}{\includegraphics{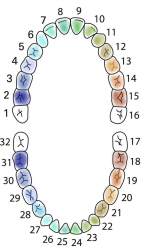}
    \label{fig:tooth}
    }} \\
    \subfigure[]{\scalebox{1}{
    \begin{tikzpicture}[
    roundnode/.style={circle, draw=black, minimum size=6mm},
    nonode/.style={circle, draw=white, thick, minimum size=6mm},
    squarednode/.style={rectangle, draw=black, thick, minimum size=9mm}]
    \node[roundnode] (T3) {3};
    \node[roundnode] (T2) [left=of T3] {2};
    \node[roundnode] (T1) [left=of T2] {1};
    \node[nonode] (T4) [right=of T3] {};
    \node[nonode] (T5) [right=of T4] {};
    \node[roundnode] (T6) [right=of T5] {12};
    \node[roundnode] (T7) [right=of T6] {13};
    \node[roundnode] (T8) [right=of T7] {14};

    \draw[solid] (T1.east) -- (T2.west);
    \draw[solid] (T2.east) -- (T3.west);
    \draw[solid] (T3.east) -- (T4.west);
    \draw[dashed] (T4.east) -- (T5.west);
    \draw[solid] (T5.east) -- (T6.west);
    \draw[solid] (T6.east) -- (T7.west);
    \draw[solid] (T7.east) -- (T8.west);
    
    \end{tikzpicture}
    \label{fig:tooth_graph}
    }}\\
    \subfigure[]{\scalebox{1}{
    \begin{tikzpicture}[
    roundnode/.style={circle, draw=black, thick, minimum size=8mm},
    squarednode/.style={rectangle, draw=black, thick, minimum size=8mm}]
    \node[squarednode] (State1) {1};
    \node[squarednode] (State0) [left=of State1] {0};
    \node[squarednode] (State2) [right=of State1] {2};
    \node[squarednode] (State3) [right=of State2] {3};

    \draw[-stealth, line width=1pt] (State0.east) -- (State1.west);
    \draw[-stealth, line width=1pt] (State1.east) -- (State2.west);
    \draw[-stealth, line width=1pt] (State2.east) -- (State3.west);
    \end{tikzpicture}
    \label{fig:MSM}
    }}
    \caption{(a) Intraoral spatial referencing of teeth: third molars $(1,16,17,32)$ are excluded, $2-16$ (upper jaw) and $18-31$ (lower jaw) are renumbered as $1-14$ and $15-28$ respectively, teeth adjacent in each jaw share the same random effect as depicted by the same colors; (b) Undirected graph connecting neighboring teeth in the upper jaw; (c) Progressive 4-state model for the GAAD data with state 0: slight PD, 1: moderate PD, 2: severe PD, and 3: missing tooth due to PD.}
    \label{fig:intro}
\end{center}
\end{figure}
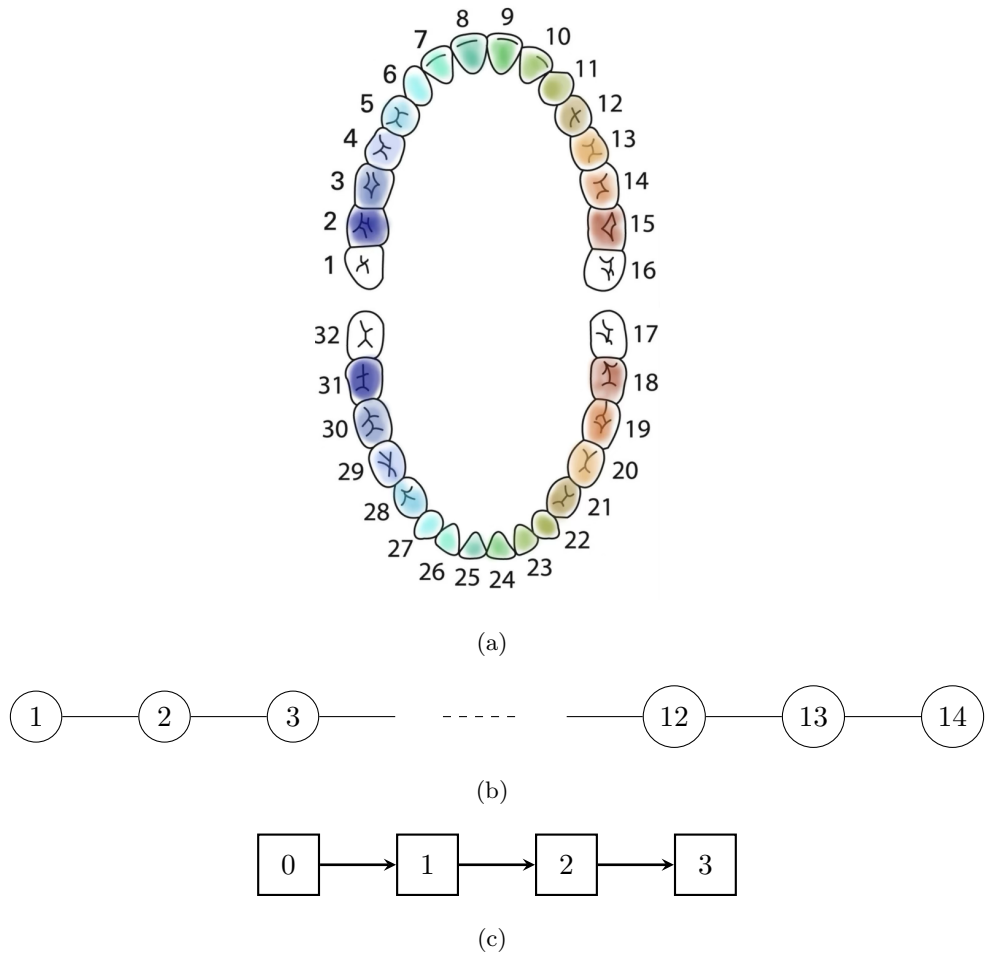

\noindent  With the objective of quantifying the association of covariates on the latent time to a missing tooth $T_{ij}$ $\big( = T_{ij}^{\,(K)} \big)$, we propose a semiparametric AFT model,
\begin{equation}
\label{eq:AFT}
\log T_{ij} = g(\ux_{ij}^\top \ubeta) + b_{ij} + \epsilon_{ij},    
\end{equation}
where $\ubeta$ is the $p$--dimensional vector of regression parameters, $g$ is an unknown monotone link function, $b_{ij}$ is a random effect incorporating the spatial referencing of teeth, and $\epsilon_{ij}$ is the random error associated with each response. 
$\epsilon_{ij}$ is assumed to have an unknown baseline distribution $F$ with corresponding density $f$.
The systematic part of $\log T_{ij}$ is modeled through $g(\ux_{ij}^\top \ubeta)$, via a monotone SIM, which provides a simplified framework for understanding a complex, nonlinear relationship between a response variable and its $p$-dimensional covariate vector, thus offering a pragmatic compromise between restrictive parametric and fully nonparametric formulations. 
$g(\ux_{ij}^\top \ubeta)$  captures how the multi-dimensional covariate, $\ux_{ij}$  influences the scalar response, $\log T_{ij}$,  beyond the baseline effect, by projecting the covariates onto a lower-dimensional space through a single index, $u_{ij} = \ux_{ij}^\top \ubeta$. The unknown link function, $g$, implicitly captures interactions and higher-order effects among predictors.
Monotonicity of the link function \citep{balabdaoui2019msim} ensures straightforward clinical interpretation of the scalar index, $u_{ij}$, and the index vector, $\ubeta$, on the response, appealing naturally to  biomedical applications. Without loss of generality, if $g$ is a monotone non-decreasing function, the expected value of the response will increase or remain the same if the index increases. If $\beta_l$, the coefficient corresponding to covariate $x_{ij,l}$, is positive, then increasing the value of $x_{ij,l}$ (with other covariates fixed) will result in a higher value of the index, thereby increasing the expected value of the response. Moreover, coefficients having a higher magnitude $|\beta_l|$ will have a stronger effect on the expected response. Furthermore, $\exp(g({\beta}_l) - {g}(0))$ represents the multiplicative effect of unit increase or change in category from $0$ to $1$ of $x_{ij,l}$ on the expected survival time, holding all other covariates fixed.
For $(g, \ubeta)$ to be jointly identifiable, $\ubeta$ is constrained to have unit norm. We further re-scale the covariates to have $\|\ux_{ij}\| \leq 1$, which ensures the support of $g$ to be $[-1,1]$, and impose $g(-1) = 0$ to have the intercept term remain implicit through the unrestricted mean of the baseline density, $f$. The constraints on $\ux_{ij}$ and $\ubeta$ aid modeling and identifiability, while the flexibility of $g$ captures varying covariate effects on the response. Additionally, we set $b_{ij^\prime} = b_{ij}$ for $j' = m-j+1$, $j = 1,2,\ldots, m/2$, under the assumption that adjacent teeth in opposite jaws of the oral cavity share the same effect (see Figure \ref{fig:tooth}), and let $\ub_i = (b_{i1}, \ldots, b_{i\,m/2})^\top \sim \mathcal{N}(0, \Sigma_b)$. 

We define the relative increment of time to the $k$th state as 
\begin{equation}
\label{eq:relative_inc}
    R_{ij}^{\,(k)} = \frac{T_{ij}^{\,(k)} - T_{ij}^{\,(k-1)}}{T_{ij}}, \quad \quad k = 1,2,\ldots, K.
\end{equation} 
following a Dirichlet distribution, such that $\uR_{ij} = \left(R_{ij}^{(1)}, \ldots, R_{ij}^{(K)} \right) \sim \text{Dir} (\alpha_1, \ldots, \alpha_K)$. Letting $R_{ij}^{\,(0)} = 0$, $R_{ij}^{\,(K+1)} = \infty$, and using (\ref{eq:states}) and (\ref{eq:relative_inc}), the observed states at their censoring times can be expressed as 
\begin{equation}
\label{eq:observed_states}
S_{ij}(C_{ij}) = k \quad \textrm{if}\quad  {T_{ij}} \sum_{l=0}^k  {R_{ij}^{\,(l)}} \leq   {C_{ij} } <  {T_{ij}} \sum_{l=0}^{k+1}  {R_{ij}^{\,(l)}}, \quad k = 0,1,\ldots, K.
\end{equation} 
Combining (\ref{eq:AFT}), (\ref{eq:relative_inc}), and (\ref{eq:observed_states}) completes our model specification. For our multistate process $\{S_{ij}(t): t \geq 0\}$ having finite state space $\mathcal{K}$, the state occupation probability (SOP) of a disease state $k$ at a given time $t \geq 0$, denoted by $ p_{ij, \,k}(t) = P \left(S_{ij}(t) = k \right)$, is defined as the probability that tooth $j$ of subject $i$ is in state $k$. We define the  transition probability (TP), denoted by $p_{ij,\,rs}(u,t) = P\left(S_{ij}(t+u) = s \mid S_{ij}(u) = r \right)$, $r\leq s$, $r,s \in \mathcal{K}$ and $t, u \geq 0$, as the probability that tooth $j$ of subject $i$ transitions to a state $s$ at time $t + u$ given that the tooth is in state $r$ at time $u$.

\subsection{Identifiability of parameters}
\label{s:identifiability}
In the following, we establish the identifiability of the fixed effects parameters in our model, in the presence of (spatial) random-effects. Previous research, often restricted to the iid setup, proved parameter identifiability \citep{linkulasekera} in a SIM for non-constant and continuous functions $g$, under the assumptions that $\| \ubeta \| =1$, and the first non-zero element of $\ubeta$ being positive. For our monotone $g$, a relaxed assumption of only $\| \ubeta \| =1$ suffices. 
To that end, identifiability of the variance components of our model is first established in Proposition \ref{p:variance_random_effects}, followed by the proof of identifiability for the fixed effects in Proposition \ref{p:identifiability}. The proofs can be found in Section \ref{s:proofs} of the Supplement.

Defining $\uy_i = (\log T_{i1}, \ldots, \log T_{im})^\top$, $\ug_i = (g(\ux_{i1}^\top \ubeta), \ldots, g(\ux_{im}^\top \ubeta))^\top$, $\uepsilon_i = (\epsilon_{i1}, \ldots, \epsilon_{im})^\top$, \eqref{eq:AFT} can be concisely written as
\begin{equation}
\label{eq:mat_AFT}
    \uy_i = \uZ \ub_i + \ug_i + \uepsilon_i, \quad\quad i = 1,2, \ldots, n
\end{equation}
where $\uZ = [\mathbb{I}_{m/2}\ \Tilde{\mathbb{I}}_{m/2}]^\top$, and $\mathbb{I}_{m/2}$, $\Tilde{\mathbb{I}}_{m/2}$ denote $m/2 \times m/2$ identity matrix and anti-diagonal matrix with unit entries, respectively. Let $ \sigma_\epsilon^2 > 0$ be the common error variance and $\Sigma_y = \uZ \Sigma_b \uZ^\top + \sigma_\epsilon^2 \,\mathbb{I}_m$ be the covariance matrix of $\uy_i$. 
Further, let $\mathbb{S}_+^{m}$ denote the space of symmetric and positive definite $m \times m$ matrices, $\mathcal{C}_\mathcal{M}$ denote the space of  continuous monotone functions on $[-1,1]$ with $g(-1)=0$, and $\mathcal{S}_{p-1}$ denote the unit sphere of dimension $p$.

\begin{proposition}
\label{p:variance_random_effects}
    The variance components of the model, $\Sigma = \big(\Sigma_b, \sigma_\epsilon^2 \big)$ in the space $\mathcal{V} = \big\{ \big(\Sigma_b, \sigma_\epsilon^2 \big): \Sigma_b \in \mathbb{S}_+^{m/2}, \, \sigma_\epsilon^2 \in (0, \infty) \big\}$, are identifiable.
\end{proposition}

\begin{proposition}
\label{p:identifiability}
Given a fixed baseline distribution $F$ of the errors and a fixed covariance matrix $\Sigma_b$ of the random effects, the fixed effects of the model, $\theta = (g, \ubeta)$ in the space $\Theta = \{(g, \ubeta): g \in \mathcal{C}_\mathcal{M}, \ubeta \in \mathcal{S}_{p-1}\}$, are identifiable.
\end{proposition}

\subsection{Bayesian inference}
\label{s:Bayes_inf}
In this section, we develop the Bayesian inferential framework for our proposed model by defining suitable priors for the model parameters. Posterior samples are simulated through the implementation of Markov chain Monte Carlo (MCMC) methods. We elaborate on the choice of priors in the following, while deferring detailed derivations of full conditionals and posterior sampling strategies to Section \ref{s:posteriors} of the Supplement. 

\subsubsection{Prior for the regression parameters}
\label{s:prior_beta}

Since $\ubeta$ is constrained to have unit norm, we put a normalized Gaussian prior on $\ubeta$ \ie let  $\widetilde \ubeta \sim \mathcal{N}(0, \sigma_\beta^2 \mathbb I_p)$ and set $\ubeta = \widetilde \ubeta / \|\widetilde \ubeta\|$. Posterior sampling is done using an efficient elliptical slice sampler \citep{murray2010elliptical}, particularly suited for performing inference in models with multivariate Gaussian priors.

\subsubsection{Model and prior for the monotone link function}
\label{s:prior_link}
A general approach to Bayesian monotone function estimation is to expand the unknown function in a basis and translate the functional constraints to linear constraints in the coefficient space. Most widely used examples include Bernstein polynomials \citep[BP;][]{Wang2012Bernstein} and monotone splines \citep{meyer2011bayesian} as the basis choices. Monotonicity of a function 
is enforced in a BP basis expansion by ordering the coefficients, but this condition is only sufficient and not necessary. There may exist monotone functions for which the BP basis expansion results in unordered coefficients, and can lead to poor approximation (see Figures \ref{fig:probit1}, and \ref{fig:probit2}).
Alternatively, higher order monotone splines are used for capturing the underlying smoothness of a function \citep{wang2008isotonic}.

In this paper, we consider first order monotone splines to expand the function $g$ in a basis and enforce smoothness through an equivalently constrained Gaussian process (GP) prior on the basis coefficients, as in \cite{maatouk2017gaussian}. 
Adapting the notation of \cite{maatouk2017gaussian}, we begin with defining a regular grid of $L+1$ equi-spaced knot points, $u_l = \delta_L(l - L/2)$, with $\delta_L = 2/L$, $l = 0,1,\ldots, L$, in the closed interval $[-1,1]$. Further, the interpolation basis functions $\{h_l\}$ at each knot point are defined as,
\begin{equation*}
    h_l(x) = h \left(\frac{x-u_l}{\delta_L} \right), \quad l = 0,1,\ldots, L,
\end{equation*}
where $h(x) = (1 - |x|)\, \mathds{1}_{[-1,1]}(x)$. The basis functions $\{h_l\}$ can be used to approximate any continuous function $g : [-1,1] \rightarrow \mathbb{R}$ by linearly interpolating the value of $g$ at the knot points, i.e., $g(x) \approx \sum_{l=0}^L g(u_l)\,h_l(x)$. Since our link function, $g$, is monotone, we bypass using $\{h_l\}$ to approximate $g$, and instead use the smoothed integrated basis $\{\psi_l\}$ defined as 
\begin{equation*}
     \psi_l(x) = \int_{-1}^x h_l(t)dt, \quad l = 0,1,\ldots L.
\end{equation*}

To that end, using the fundamental theorem of calculus for any continuously differentiable function $g$ on $[-1,1]$ we have
\begin{equation}
 \label{eq:expansion_g}
     g(x) = g(-1) + \int_{-1}^x g^\prime(t)\, dt.
\end{equation}
Approximating $g^\prime$ in (\ref{eq:expansion_g}) using the interpolation basis, $\{h_l\}$, and enforcing $g(-1)=0$, we have
\begin{equation}
 \label{eq:basis_expansion}
     g(x) \approx g(-1) + \sum_{l=0}^L g^\prime(u_j) \int_{-1}^x h_l(t)\, dt \ = \sum_{l=0}^L \xi_{l} \,\psi_l(x).
\end{equation}
\cite{maatouk2017gaussian} proved that the function $g$ is monotone non-decreasing if and only if $\xi_l \geq 0$ for all $l = 0,1,\ldots,L$. Thus, if $g$ is assumed to be monotone non-decreasing, this constraint can be equivalently expressed by restricting the basis coefficients $\uxi = (\xi_0, \ldots, \xi_{L})^\top$ to lie in the positive orthant, 
\begin{equation}
\label{eq:constaints}
    \mathcal{C}_\xi = \{\uxi \in \mathbb{R}^{L+1} : \xi_l \geq 0, \ l = 0, 1, \ldots, L\}.
\end{equation}
A Gaussian process \citep[GP;][]{rasmussen2006gaussian} prior on $g$ induces a Gaussian prior on $\uxi$, since derivatives of a GP having a sufficiently smooth covariance kernel are again a GP. For monotone non-decreasing functions, \cite{maatouk2017gaussian} proposed a Gaussian prior on $\uxi$ by computing the induced prior on $\left(g^\prime(u_0), \ldots, g^\prime(u_L)\right)$ from a GP prior on $g$ and restricting $\uxi$ to lie in $\mathcal{C}_\xi$, defined in (\ref{eq:constaints}). This induced prior on $\xi$ from a stationary GP prior on $g$ is no longer stationary. In addition to $\mathcal{O}(L^3)$ complexity involved in sampling from truncated multivariate Gaussians, this specification requires $\mathcal{O}(L^2)$ computation for inverting and storing the covariance matrix, posing challenges to the sampling algorithm for large $L$.
Deviating slightly from the formulation of \cite{maatouk2017gaussian} and following the specification of \cite{Ray2020EfficientBS}, we put a stationary GP on the first derivative of $g$, 
\begin{equation}
\label{eq:xi_prior}
    p(\uxi) \propto \mathcal{N}(0, \tau^2 K) \ \mathds{1}_{\mathcal{C}_\xi}(\uxi),
\end{equation}
where $\tau^2 > 0$ controls the prior variance and is assigned a non-informative inverse-gamma $\mathcal{IG}(a_\xi, b_\xi)$ prior, and $(K)_{ij} = k(u_i - u_j)$, $k(\cdot)$ is the stationary Matérn covariance kernel with smoothness $\nu > 0$ and length-scale $\ell > 0$. 
This specification allows the covariance matrix of $\uxi$ to have a Toeplitz structure, enabling us to use the circulant embedding techniques by \cite{wood1994simulation} that employ discrete fast Fourier transforms. An elliptical slice sampler \citep{murray2010elliptical} is used to sample $\uxi$ by approximating the indicator function in (\ref{eq:xi_prior}) with a smooth approximant, $\mathds{1}_{\mathcal{C}_\xi}(\uxi) \approx \mathbb{J}_\eta(\uxi) = \prod_{l=0}^{L} (1+e^{\, - \eta\,\xi_l})^{-1}$,
where $\eta > 0$ controls the quality of the approximation, with a higher value giving a better approximation.  This algorithm draws a sample from $\mathcal{N} (0, \tau^2 K)$ with $\mathcal{O}(L \log L)$ computations instead of $\mathcal{O}(L^3)$, enabling substantial reduction in complexity.

\subsubsection{Prior for the covariance matrix of the spatial random effects}
\label{s:prior_spatial_random_effects}

The undirected graph $G$ in Figure \ref{fig:tooth_graph} connects neighboring teeth in the upper jaw. It suffices to focus on one jaw, as each tooth in the upper jaw shares the same random effect with its corresponding tooth in the lower jaw. Let $W$ and $E_W$ respectively denote the adjacency matrix and a diagonal matrix with the $j$th diagonal entry representing the number of neighbors at location $j$ of $G$. We impose a conjugate inverse-Wishart prior on $\Sigma_b$, $\Sigma_b \sim \mathcal{IW}(c, S)$ with degrees of freedom $c = m/2 + 2$ and scale parameter $S = (E_W - \rho W)^{-1}$, where $\rho$ controls the degree of spatial dependence. The prior is thereby loosely centered around a conditionally auto-regressive model \citep[CAR;][]{besag1974spatial} that encourages strong association between neighboring teeth.

\subsubsection{Prior for the density of the errors}
\label{s:prior_errors}
To estimate the unknown baseline density $f$, we assign a Dirichlet process (DP) mixture prior of Gaussians.
The Dirichlet process \citep[DP;][]{ferguson1973bayesian}, $DP(\gamma, G_0)$, is a probability measure on probability measures, where $\gamma > 0$ is the concentration parameter and $G_0$ is the base probability measure. A DP mixture prior of Gaussians on $f$ posits that 
\begin{align*}
    \epsilon_{ij} \mid \mu_{ij}, \sigma^2_{ij} &\sim \mathcal{N}(\mu_{ij}, \sigma^2_{ij}), & (\mu_{ij}, \sigma^2_{ij}) \mid G &\sim G, & G &\sim DP(\gamma, G_0) .
\end{align*}
The stick-breaking construction by \cite{sethuraman1994constructive} allows us to express $G$ as 
\begin{align*}
    G(\mu, \sigma^2) & = \sum_{h=1}^\infty \pi_h \delta_{(\phi_h, \, s^2_h)}(\mu, \sigma^2), & 
    \left(\pi_h\right)_{h=1}^\infty &\sim \text{GEM}(\gamma), & \left(\phi_h, s^2_h \right)_{h=1}^\infty &\sim G_0, 
\end{align*}
where GEM stands for Griffiths, Engen and McCloskey (see \cite{pitman2002poissondirichlet} for details) and $\delta_{(\phi,\, s^2)}(\cdot)$ is the Dirac measure at $(\phi, s^2)$. 

We choose a conjugate normal inverse-gamma $\mathcal{NIG}(\mu_{\epsilon}, \nu_{\epsilon}, a_{\epsilon}, \lambda_{\epsilon})$ as the base distribution $G_0$ and put a gamma prior $\mathcal{G}(a_\gamma, \lambda_\gamma)$ on $\gamma$, in the light of \cite{escobar1995bayesian}.
Posterior sampling is carried out using a blocked Gibbs sampler \citep{ishwaran2001gibbs} that uses a finite truncation for the number of mixture components upto a large number $H$, and is known to exhibit scalability and good mixing.

\subsubsection{Prior for the Dirichlet parameters of the relative disease increment times} 
\label{s:prior_dirichlet}
We impose independent gamma priors on $\mathcal{G}(a_\alpha, \lambda_\alpha)$ on $\alpha_k$, $k = 1, 2, \ldots, K$, and perform posterior sampling using a random walk Metropolis within Gibbs algorithm, tuned for an acceptance ratio of $0.2 - 0.4$.

\subsection{Estimation of state occupation and transition probabilities}
\label{s:probabilities}

The SOP and TP for each disease state (along with their credible intervals) can be estimated using Monte Carlo posterior samples. We will employ the joint distribution of $\uR_{ij}$ and
the marginal distribution of $\log T_{ij}$, which can be specified as a mixture of Gaussians using \eqref{eq:mat_AFT},
\begin{equation}
\label{eq:dist_logT}
    \log T_{ij} \sim \sum_{h=1}^\infty \pi_h \, \mathcal{N}\left(\, g(\ux_{ij}^\top \,\ubeta) + \phi_h, \ \sigma^2_{b,j} + s^2_h \,\right)
\end{equation}
where $\sigma^2_{b,j} = \uz_j^\top \Sigma_b\, \uz_j$, \,  $\uz_j^\top$ denoting the $j$th row of $\uZ$.
Corresponding to each posterior sample of our model parameters and each tooth of each subject, we draw $B$ independent samples from the joint distribution of $\uR$ and that from the marginal distribution of $T$ using \eqref{eq:dist_logT}, details of which are outlined in Section \ref{s:SOP_TP_supp} of the Supplement. Henceforth, we drop the subscripts $ij$ for clarity of notation. Let $\{T_{i^\prime}, \uR_{i^\prime} : i^\prime = 1, 2, \ldots, B\}$ denote the Monte Carlo samples. 
Letting $R^{(0)}_{i^\prime} = 0$, $R^{(K+1)}_{i^\prime} = \infty$, we define indicator functions $\eta_k(t\,;\, T, \uR) = \mathds{1} \big( T \sum_{l=0}^k  {R^{\,(l)}} \leq   t <  T \sum_{l=0}^{k+1}  R^{\,(l)} \big)$ corresponding to the SOPs using \eqref{eq:states} and \eqref{eq:observed_states} for each state $k \in \mathcal{K}$, 
and estimate the probabilities with their Monte Carlo averages, $\widehat{p}_k(t) = (1/B) \ \sum_{i^\prime = 1}^{B} \eta_k(t \,;\,T_{i^\prime}, \uR_{i^\prime})$. The TPs defined as a conditional probability can similarly be estimated by the Monte Carlo method, $\widehat{p}_{rs}(u,t) = {\sum_{i^\prime = 1}^B \eta_s(t + u \,;\,T_{i^\prime}, \uR_{i^\prime}) \ \eta_r(u \,;\,T_{i^\prime}, \uR_{i^\prime})} \bigg/ {\sum_{i^\prime = 1}^B \eta_r(u \,;\,T_{i^\prime}, \uR_{i^\prime})}$, $r \leq s$, $r, s \in \mathcal{K}$.

\section{Application: GAAD data} 
\label{s:data_analysis}
In this section, we illustrate our methodology via application to the GAAD dataset. Progression of PD is usually assessed via clinical attachment level \citep[CAL;][]{bandyopadhyay2016non}, defined as the distance (in mm whole numbers) down a tooth’s root that is no longer attached to the surrounding bone by the periodontal ligament, and recorded at six pre-specified tooth sites for each of the $28$ teeth, excluding the four third molars. Following the definition \citep{Wiebe2000ThePD} of \textit{chronic periodontitis severity}, our tooth-level PD status comprises average CAL (avCAL) at the corresponding tooth-sites, and categorized into $4$ states: $0$ denotes slight PD (avCAL $\in [0,2)$ mm), $1$ denotes moderate PD (avCAL $\in [2,5)$ mm), $2$ denotes severe PD (avCAL $\geq 5$ mm) and $3$ denotes missing tooth (avCAL is unavailable, due to past PD). The natural history of chronic PD can be viewed as a staged progressive disease process, from \textit{gingivitis} to \textit{periodontitis} characterized by irreversible pathological changes \citep{mdala2014comparing}, as shown in Figure \ref{fig:MSM}. In cross-sectional studies like GAAD, the true inspection time for each tooth should be calculated from the adult tooth emergence time, which is typically unavailable. In lieu of this, we approximate the tooth inspection time for all subjects by taking the difference between the subject’s age (recorded at clinic visit) and the permanent dentition times of US adults published by the \cite{adaeruptioncharts}.  The data includes $n = 288$ subjects with $m = 28$ teeth for each subject, $46\%$ of which are in state $0$, $19\%$ in state $1$, $2\%$ in state 2 and the remaining $32\%$ in state $3$. Related subject-level covariates in the data include gender ($1$ = female; $0$ = male), body mass index (BMI; in $\text{kg}/\text{m}^2$), smoking status ($1$ = past/current smoker, $0$ otherwise) and glycosylated hemoglobin (HbA1c; $1$ = high/uncontrolled, $0$ = controlled). We also considered a binary tooth-level covariate `Maxilla' ($1$, if tooth is in upper jaw, $0$, if tooth in lower jaw/mandible).  The subjects have a mean age of $55$ years, ranging from 26 to 87 years. Around $76\%$ of the subjects are female, $68\%$ are categorized as obese (with a BMI $\ge$ 30), $31\%$ are smokers, and $59\%$ have uncontrolled HbA1c levels. BMI, being a continuous covariate is standardized to have mean $0$ and variance $1$.

We considered six competing models during data fitting, which differ in terms of their specifications for the monotone link basis, random effects, and distribution of random errors:
\begin{enumerate}
    \item S-GP-DP: Spatial random effects with constrained GP basis for the link function, and DP mixture prior on the error density -- our proposed model.
    \item S-BP-DP: Spatial random effects with Bernstein polynomial (BP) basis for the link function and DP mixture prior on the error density.
    \item S-GP-N: Spatial random effects with constrained GP basis for the link function, and normal density for errors.
    \item S-BP-N: Spatial random effects with BP basis for the link function, and normal density for errors.
    \item S-lin-DP: Linear model (identity link) with spatial random effects, and DP mixture prior on the error density.
    \item NS-GP-DP: Non-spatial random effects at the subject-level ($b_{i1} = \ldots = b_{im} = b_i$) with constrained GP basis for the link function, and DP mixture prior on the error density.
\end{enumerate}

For our analysis, we set $\sigma_\beta^2 = 100$, though any arbitrary positive value can be used since $\Tilde{\beta}$ is scaled by its norm. For the link function, we use default choices of $\nu = 0.75$, with $\ell$ chosen such that the correlation at a maximum covariate separation is $0.05$ \citep{Ray2020EfficientBS}, and set $\eta = 100, a_\xi = 0.01$ and $b_\xi = 0.01$. The degree of spatial dependence is set to $\rho = 0.9$ in the prior on $\Sigma_b$. For the DP mixture, we set $H=10$, $\mu_\epsilon = 0$, $\nu_\epsilon = 1$, and choose $a_\epsilon$, $b_\epsilon$ to yield a prior mean and variance of $s_h^2$ as $0.1$ and $10$, respectively. The parameters in the gamma prior on $\gamma$ is chosen to have mean $1$ and variance $10$. Since information on the state transition times from the observed data is weak, we use a weakly informative gamma prior on $\ualpha$ with mean $1$ and variance $100$. More details on the choice hyperparameters for each prior are provided in Section \ref{s:hyperparameters} of the Supplement.

We generated posterior MCMC samples of size $\mbox{70,000}$ iterations with a burn-in of $\mbox{20,000}$ and collected every 50th sample to thin the chain. 
Posterior convergence was assessed using trace plots and the Gelman-Rubin statistic $\hat{R}$. For model selection tools, we considered the Watanabe-Akaike information criterion (WAIC), and the leave-one-out cross-validation (LOO-IC) using Pareto-smoothed importance sampling 
available in \texttt{R} package \texttt{loo}, and enlist them in Table \ref{t:model_selection}. Models with smaller WAIC and LOO-IC values suggest a better fit to the observed data. For all models except (5), the optimal number of basis functions $L$ was set to $30$, chosen by minimizing WAIC as $L$ varied from 10 to 45 in increments of 5; details are provided in Section \ref{s:knots} of the Supplement.

\begin{table}[ht]
\caption{Model selection criteria for comparing the six competing models on the GAAD data.}
\label{t:model_selection}
\begin{center}
\begin{tabular}{lcccccc}
\hline
 & S-GP-DP & S-BP-DP & S-GP-N & S-BP-N & S-lin-DP & NS-GP-DP \\ 
  \hline
WAIC & 11189.0 & 11567.6 & 12502.2 & 12523.4 & 11779.2 & 13836.0 \\ 
LOO-IC & 11282.3 & 11604.7 & 12488.8 & 12502.1 & 11851.0 & 13874.6 \\ 
\hline
\end{tabular}
\end{center}
\end{table}

The non-spatial model having the highest value of both WAIC and LOO-IC validates the necessity of incorporating the intraoral spatial dependencies between the teeth. Among the spatial models, the 3 models having a DP mixture prior on the error density have lower WAIC and LOO-IC, warranting the improved flexibility of DP over normal errors. This is further supported by the larger variability observed in the boxplots of model-based residuals for the models with a normal errors, compared to the DP errors; see Section \ref{s:residuals} of the Supplement. The models with constrained GP basis for the monotone link show better predictive performance compared to the corresponding ones with BP basis. Clearly, our proposed S-GP-DP model outperforms all its competitors.

\subsection{Estimation of fixed effects and spatial dependence}
\label{s:fixedeffects}

\begin{figure}
\begin{center}
    \subfigure[Estimated regression coefficients, $\widehat{\ubeta}$.]{\scalebox{0.7}{\includegraphics{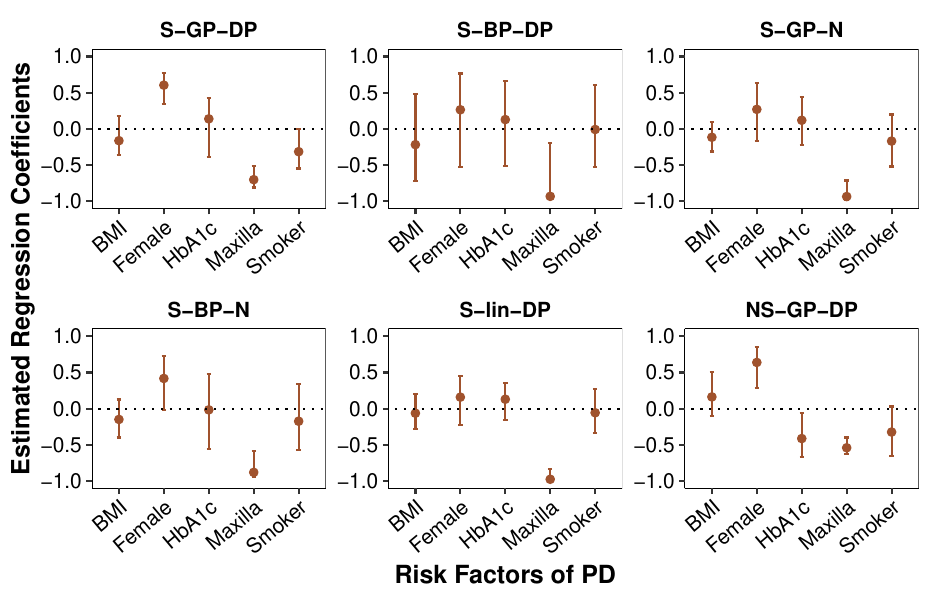}}
    \label{fig:reg}
    }
    \subfigure[{Estimated monotone link function, $\hat{g}(u)$ against the single index, $u \in [-1,1]$.}]
    {\scalebox{0.68}{\includegraphics{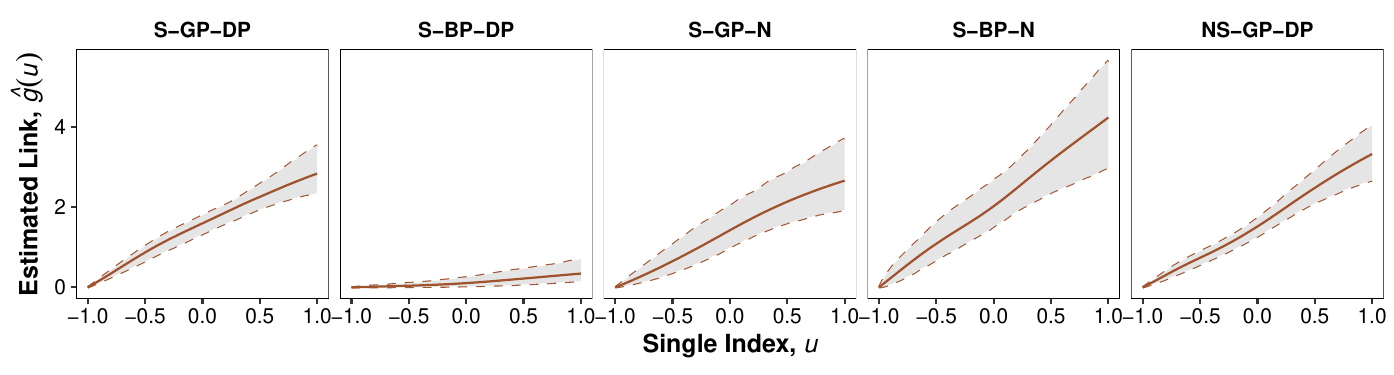}}
    \label{fig:link}}
    \subfigure[Estimated precision matrix $\Omega_b$ of the $14$ tooth-level random effects shared between the upper and lower jaws.]{\scalebox{0.54}{\includegraphics{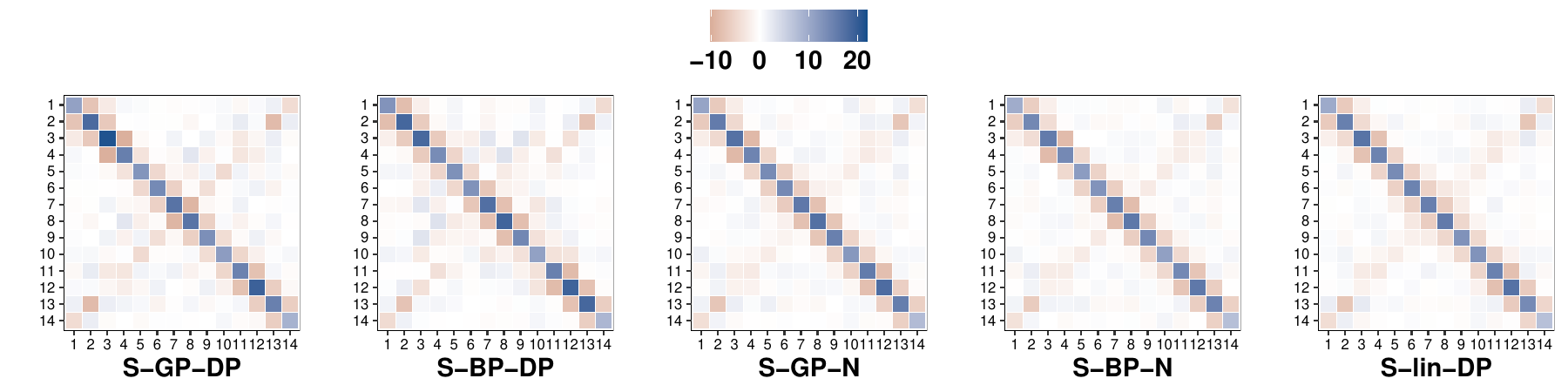}}
    \label{fig:precision_mat}}
    \caption{Posterior mean and $95\%$ credible intervals for (a) regression parameters, (b) monotone link function, and (c) heatmap of the estimated precision matrix of the spatial random effects, under the competing models on the GAAD data.}
    \label{fig:fixedeffects}
\end{center}
\end{figure}

Posterior estimates and $95\%$ credible intervals (CIs) of the regression parameters obtained from fitting the six competing models are presented in Figure \ref{fig:reg}. 
The CI for upper jaw (maxilla) excluded $0$, and the posterior estimates were negative with highest magnitude for all six models, indicating that this covariate has the strongest influence on the index. Consequently, the index and hence the survival time, for a maxillary tooth is lower than that for a mandibular tooth. This is in tune with findings revealing higher loss and extraction of maxillary tooth, than mandibular, due to PD \citep{chraibi2023reasons}. Strong evidence of a higher index is observed in females compared to males from the S-GP-DP, S-BP-N and NS-GP-DP models, resonating with higher prevalence of chronic PD in males \citep{ioannidou2017sex}. This association, however,  lacks significance (though positive) in the other models. Smoking is revealed to negatively impact the index in all models, leading to reduced survival, thereby corroborating with published literature \citep{zee2009smoking}, with significant impact in the S-GP-DP and NS-GP-DP models. BMI and glycemic levels do not seem to significantly affect the index, with the exception of a strong negative impact of HbA1c in the NS-GP-DP model. 

Figure \ref{fig:link} plots the link functions estimated from the models, along with their $95\%$ CIs. The range of the link functions estimated by the three models having constrained GP basis are similar. The spatial models with BP basis and normal errors have a higher range, while that with DP mixture errors has a very low range. This is in alignment with the performance of these models in our simulation studies (see Section  \ref{s:sims} for details). 
The estimates $\exp(\hat{g}(\hat{\beta_l}) - \hat{g}(0))$ for the significant predictors \ie gender, smoking and maxilla from our S-GP-DP model indicate females have approximately $2.2$ times the survival time of males, smokers have about $0.65$ times that of non-smokers, and maxillary teeth exhibit a survival time of $0.3$ times that of mandibular teeth. The S-GP-N and S-BP-N models show similar estimates. These estimates from the S-BP-DP model are close to $1$ for all predictors, as evident from the low range of its estimated link function. The NS-GP-DP model additionally identifies glycemic level as a significant predictor, showing that a high glycemic level reduces survival time to 0.56 times that of a low level.

Heatmaps of the posterior estimate of the precision matrix $\Omega_b$ of the random effects in the five spatial models are illustrated in Figure \ref{fig:precision_mat}. Strong correlations are seen between first molars (teeth 2 and 13), which are mirror images in opposite maxillary quadrants, consistent with clinical findings which highlight that molars and premolars generally exhibit higher levels of PD \citep{kindler2018third}. The matrices are mostly concentrated along their tridiagonal, confirming the neighborhood association depicted in Figure \ref{fig:tooth_graph} and suggests that spatial dependence weakens considerably beyond the immediate neighboring tooth.

\subsection{Estimation of state occupation and transition probabilities}

\begin{figure}
\begin{center}
    \subfigure[]{\scalebox{0.57}{\includegraphics{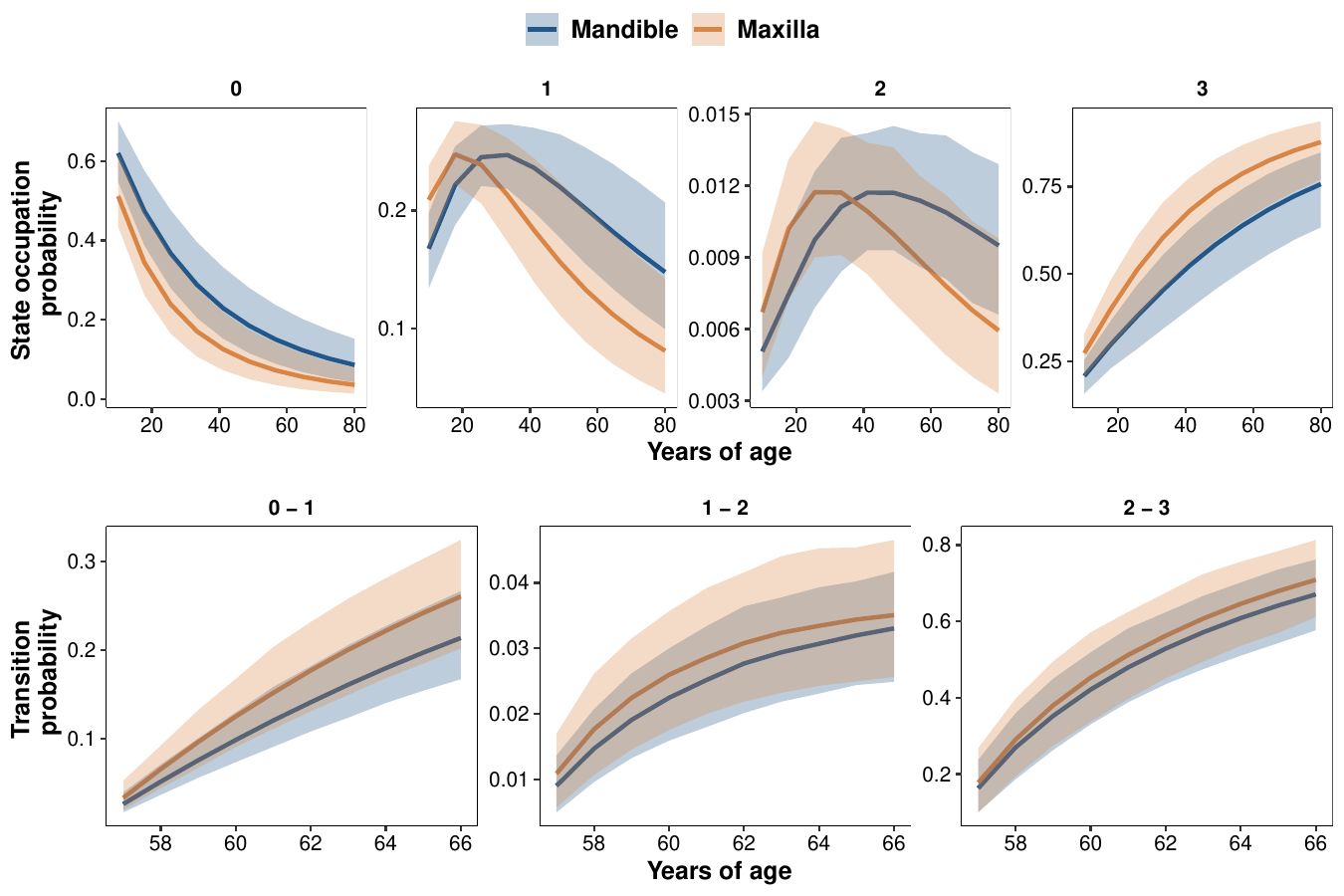}
    \label{fig:SOP_TP_jaw}}
    }
    \subfigure[]{\scalebox{0.57}{\includegraphics{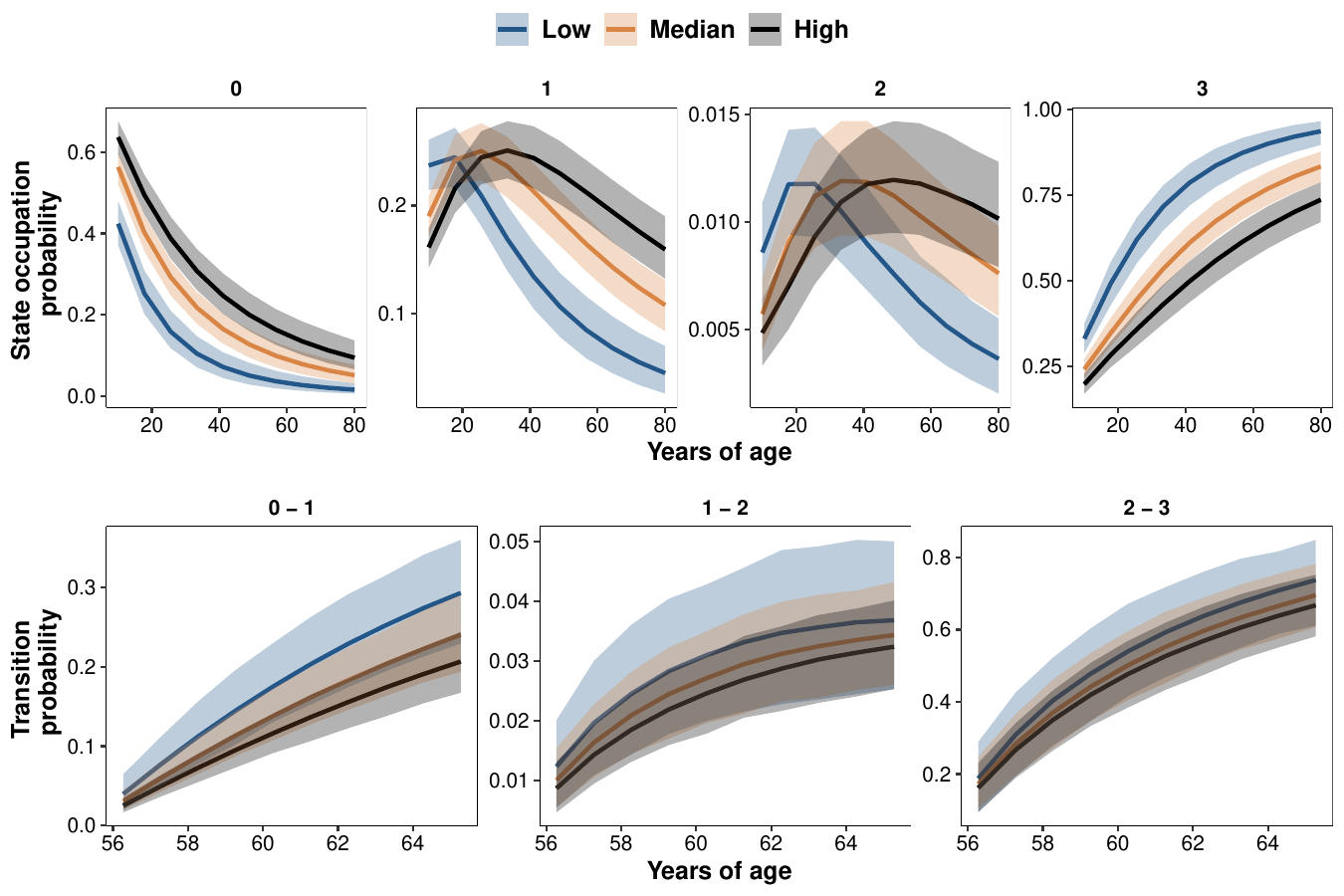}
    \label{fig:SOP_TP_SI}}
    }
    \label{fig:SOP_TP}
    \caption{Plots of the posterior mean and $95\%$ CI of the state occupation probabilities (SOPs) over the years $10-80$, and transition probabilities (TPs) for the next 10 years (upper two panels), for the maxillary and mandibular second molars of a random subject (female, smoker, age = 56 years, BMI = 35.33, with uncontrolled HbA1c), based on $\mbox{10,000}$ Monte Carlo samples. The lower two panels plot similar SOPs and TPs, now corresponding to the minimum, median and maximum values of the estimated index $\hat{u}_{ij} = \ux_{ij}^\top \hat\ubeta$, where $\hat{\ubeta}$ = posterior mean of $\ubeta$. While 0, 1, 2, and 3 denote the states, $0-1$, $1-2$, and $2-3$ denote the transitions between the states. }
\end{center}
\end{figure}

To illustrate the use of our model to probe the association between a covariate and the state occupation and transition probabilities of teeth, we arbitrarily consider a female smoker aged 56 years (median), with a BMI of 35.33 (mean) and uncontrolled HbA1c.
Figure \ref{fig:SOP_TP_jaw} plots the posterior mean and $95\%$ CI of the SOPs over the years 10--80, and TPs for the next 10 years (with the baseline age set to the median of 56 years), for the maxillary (upper jaw) and mandibular (lower jaw) second molar (teeth numbers 1 and 28, respectively) of this subject, based on $\mbox{10,000}$ Monte Carlo samples. On the overall, our model indicates that maxillary location negatively affects tooth survival. The SOPs for state $0$ decreases while that for state $3$ increase over time, with the mandibular molar having higher (lower) SOP for state 0 (3), than that of the maxillary counterpart, as one would naturally expect. Consequently, intersecting trajectories for the SOPs of states $1$ and $2$ was observed, initially showing an increasing trend for both jaws before decreasing, with the change-point occurring at a higher age for mandible than that for maxilla.

To collectively summarize the combined effect of all the covariates on these probabilities, Figure \ref{fig:SOP_TP_SI} plots the posterior mean and $95\%$ CIs of the SOPs over the years 10--80, and TPs for the next 10 years, corresponding to the minimum, median and maximum value of the estimated index $\hat{u}_{ij} = \ux_{ij}^\top \hat\ubeta$, where $\hat{\ubeta}$ denotes the posterior mean of $\ubeta$. The baseline age for the TPs is again set to 56 years (median). Quite intuitively, patterns similar to Figure \ref{fig:SOP_TP_jaw} are observed. Decreasing (increasing) trend in the SOP of states 0 (state 3) is observed, with a higher probability at any given time for a higher index, with intersecting trajectories for states 1 and 2. A lower index corresponds to an elevated chance of transitioning to a higher state of the disease at any given time. We further note that the estimated probability of occupying state $2$ and that of transitioning from state $1$ to state $2$ is quite low compared to the other SOPs and TPs respectively. This can be partially attributed to the fact that only $2\%$ of the observations in our data occupied state $2$.

\section{Simulation studies}
\label{s:sims}

In this section, we conduct simulations using synthetic data to (a) investigate the frequentist finite sample properties, and (b) the effect of model misspecification in recovering the regression parameters $\ubeta$, and the link function $g$, while using our proposed model. Henceforth, we refer to them as Simulations 1 and 2, respectively, and compare the performances of models (1)--(4) outlined in Section \ref{s:data_analysis}. 

At the onset, we outline the data generating process followed in the two simulation settings. Our data generation mimics a practical setting as observed in the GAAD study. We consider $p = 3$ covariates, the highest disease state $K = 3$, $m = 10$ teeth for every subject, and vary the sample size as $n = 50, 100, 300$ and $500$. The data generation steps are as follows: 

\bigskip
\noindent {\bf Step 1.} Let $\widetilde \ux_{ij} = (\widetilde x_{ij}^{\,(1)}, \widetilde x_{ij}^{\,(2)}, \widetilde x_{ij}^{\,(3)})$ denote the vector of covariates, for every $i$ and $j$. Here $\widetilde x_{ij}^{\,(1)}$ and $\widetilde x_{ij}^{\,(2)}$ denote continuous and discrete subject level covariates respectively, which are generated by drawing $\widetilde x_{i}^{\,(1)} \sim \text{Unif}(-5,5)$,  $\widetilde x_{i}^{\,(2)} \sim \text{Bin}(1,0.5)$
and setting $\widetilde x_{ij}^{\,(1)} = \widetilde x_{i}^{\,(1)}$ and $\widetilde x_{ij}^{\,(2)} = \widetilde x_{i}^{\,(2)}$ for all $j$. Also, $\widetilde x_{ij}^{(3)}$ takes the value $1$ for $j> m/2$ and $0$ otherwise (mimicking jaw indicator).
The re-scaled covariates are obtained as $\ux_{ij} = \widetilde \ux_{ij} / \max \| \widetilde \ux_{ij}\|$.

\noindent {\bf Step 2.} The true regression parameters are set at $\ubeta^{0} = \widetilde \ubeta^{0} \,/\, \|\widetilde \ubeta^{0} \| $ with $\widetilde \ubeta^{0} =(-1, 1, -1) $. 

\noindent {\bf Step 3.} Generate the random effects $\ub_i = (b_{i1}, \ldots, b_{i\,m/2})^\top \sim \mathcal{N}(0, \Sigma_b)$, where $\Sigma_b = 0.1^2 (E_W - 0.9W)^{-1}$ \ie from a parametric CAR density, with $E_W$ and $W$ constructed according to Section \ref{s:prior_spatial_random_effects}. Set $b_{ij^\prime} = b_{ij}$ for $j^\prime = m-j+1$, $j = 1, \ldots, m/2$, 

\noindent {\bf Step 4.} The random errors $\epsilon_{ij}$ under the two simulation settings are generated from a discrete mixture of Gaussians, and a mixture of Gaussian and $t$ densities respectively, as follows: (a) {Simulation 1}: $\epsilon_{ij} \sim (1/3)\, \mathcal{N}(-0.5, 0.1^2) + (1/3)\, \mathcal{N}(0, 0.1^2) + (1/3)\, \mathcal{N}(0.5, 0.1^2)$; (b) {Simulation 2}: $\epsilon_{ij} \sim 0.9\, \mathcal{N}(0, 0.1^2) + 0.1\, t_3 $.
Here, $t_3$ denotes a $t$ density with $3$ degrees of freedom. The  $t_3$ distribution introduces heavy tails and different error behavior than a simpler discrete mixture of Gaussians.

\noindent {\bf Step 5.} We consider the following true monotone link functions :
\begin{align*}
    g^0_1(x) &= c_1 \cdot \left\{ \Phi \left( \frac{(x+1)/2 \ - \ 0.5}{0.2}\right) - \Phi \left( { - \ 0.5}/{0.2}\right) \right\}  \mathds{1}_{[-1,1]}(x)\\
    g^0_2(x) &= c_2 \cdot \bigg\{ \left(\frac{x+1}{2}\right)^2 + \left(\frac{x+1}{2}\right)^3 \bigg\} \mathds{1}_{[-1,1]}(x)
\end{align*}
where $\Phi(\cdot)$ denotes the CDF of a standard Gaussian distribution and the constants $c_1$, $c_2$ are so chosen that the signal-to-noise ratio (SNR) is fixed at $5$. Note that $g_1^0$ is both convex and concave with an inflection point at 0, while $g_2^0$ is strictly convex. They allow us to assess the model's ability to capture diverse relationships between the predictors and the response.

\noindent {\bf Step 6.} The true time to the missing tooth (denoting the highest state) $T_{ij}$ are generated using (\ref{eq:AFT}). The relative time increments of the disease states are generated as $\uR_{ij} \sim \text{Dir}(3,4,2)$ and the random inspection times $C_{ij}$ are generated from $\mathcal G(1, 0.2)$  and  $\mathcal{G}(0.5, 0.2)$ for the two link functions, respectively. Time to the $k^{th}$ disease state are obtained using (\ref{eq:relative_inc}). The parameters of the generating distribution of $C_{ij}$ are so chosen to avoid large discrepancies between the proportion of individuals having disease status $0$ and those having status $K$.

\noindent {\bf Step 7.} The disease states at their inspection times $S_{ij}(C_{ij})$ are determined using (\ref{eq:observed_states}).

\begin{figure}
    \centering
    \includegraphics[width = 12.5cm]{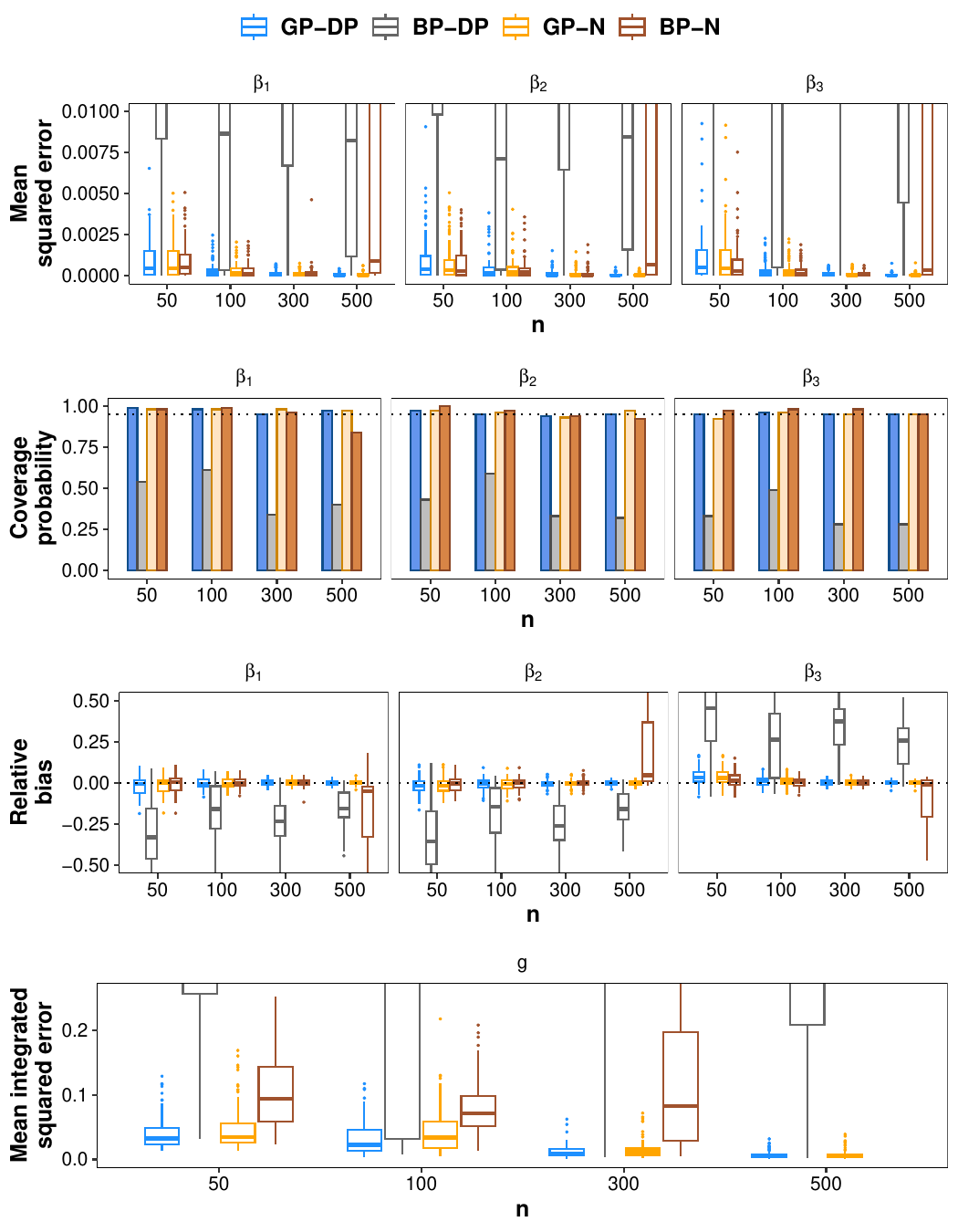}
    \caption{Barplots of coverage probabilities (CP, second), boxplots of mean squared error (MSE, topmost) and relative bias (RB, third) of estimated regression coefficients; and boxplots of mean integrated squared errors (MISE, lowermost) of estimated link function when the true link is $g_1^0$ under Simulation 1: the fitted model is well specified. Plots show variation across 100 simulation replicates.}
    \label{fig:probit1}
\end{figure}

\begin{figure}
    \centering
    \includegraphics[width = 12.5cm]{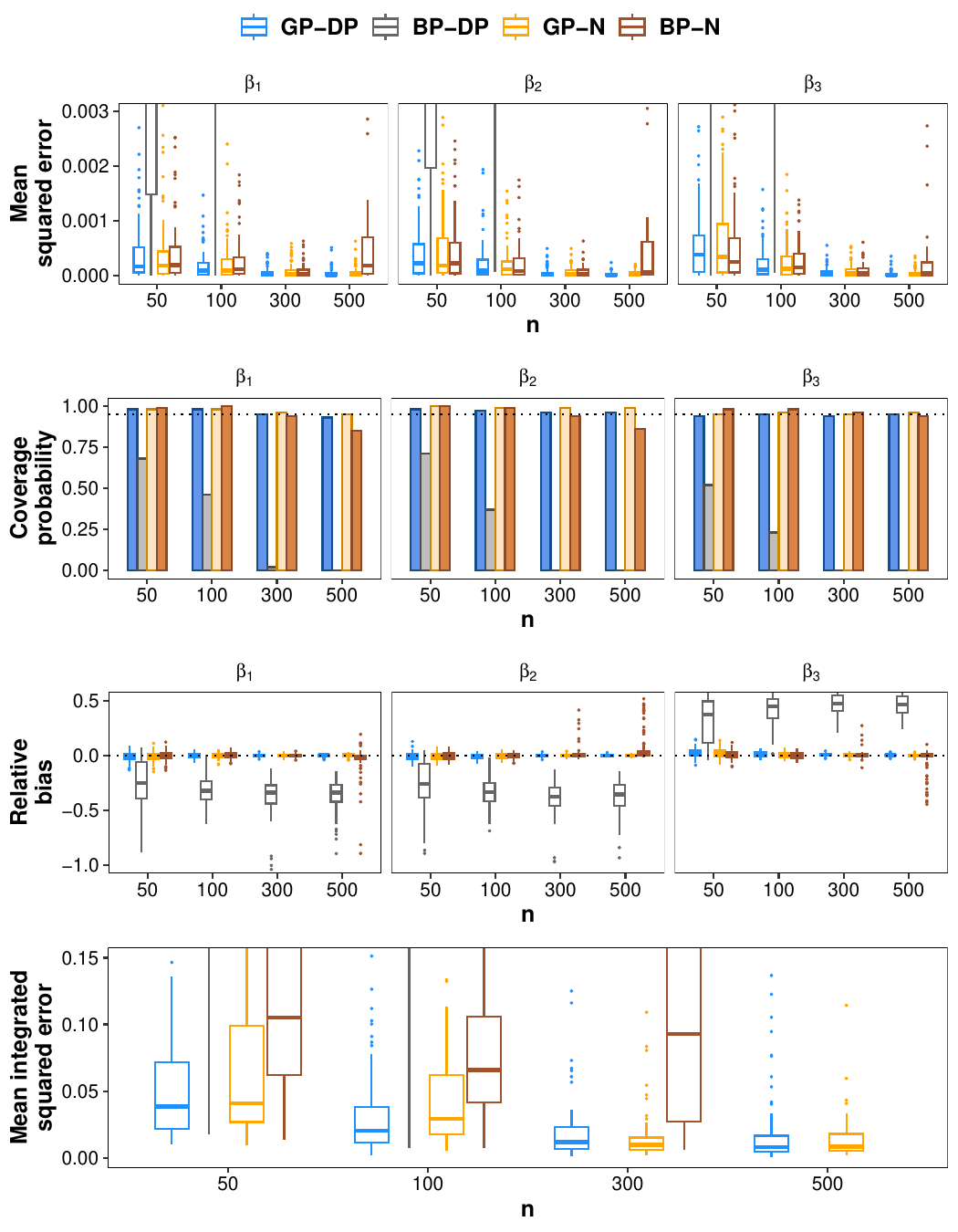}
    \caption{Barplots of coverage probabilities (CP, second), boxplots of mean squared error (MSE, topmost) and relative bias (RE, third) of estimated regression coefficients; and boxplots of mean integrated squared errors (MISE, lowermost) of estimated link function when the true link is $g_1^0$ under Simulation 2: the fitted model is misspecified. Plots show variation across 100 simulation replicates.}
    \label{fig:probit2}
\end{figure}

\medskip
For all competing models, we conduct MCMC sampling for $\mbox{7,000}$ iterations and discard the first $\mbox{5,000}$ as burn-in. Posterior estimates are summarized over $100$ simulation replicates. Estimation performance is compared using mean squared error (MSE), relative bias (RB) and coverage probability (CP) of the estimated $95\%$ CIs for each component of $\ubeta$ and mean integrated squared error (MISE) for the link functions, respectively. Let $\hat{\beta}_j$ denote the posterior mean of $\beta_j$. Component-wise MSE and RB are calculated as 
$$
\text{MSE}(\hat \beta_j) = (\hat \beta_j - \beta^{0}_j)^2, \qquad \text{RB}(\hat \beta_j) = (\hat \beta_j - \beta^{0}_j)/\beta^0_j.
$$
For estimating the two link functions $g^0_1$ and $g^0_2$, we consider $100$ equidistant grid points $\{y_1, y_2, \ldots, y_{100}\}$ in $[-1,1]$, and let $\hat g_j(y_i)$ denote the $j$th estimated link function at the $i$th grid point. The MISE of $\hat{g}_j$ is calculated as $$\text{MISE} (\hat{g}_j ) = \frac{1}{100} \sum_{i=1}^{100} \left\{ \hat{g}_j (y_i) - g_j^{\,0} (y_i)\right\}^2.$$ Results considering the true link function as $g^0_1$ are presented here, while those from $g^0_2$ are deferred to Section \ref{s:add_sim} of the Supplement. We provide boxplots of the MSE, RB and MISE and barplots of the CP across the 100 replicates over increasing $n$ under Simulations 1 and 2 in Figures \ref{fig:probit1} and \ref{fig:probit2} respectively.

\textbf{Simulation 1}: When the model is correctly specified, the S-GP-DP and S-GP-N models yield an overall superior estimation performance than the models with BP basis. Decreasing trends in the boxplots of MSE, RB and MISE with increasing sample sizes imply posterior estimates from the models with constrained GP basis are consistent. The desired $95\%$ coverage is attained by each $\hat \beta_j$ for both small and large sample sizes. While S-BP-N shows a poor performance in terms of all aforementioned metrics, the S-BP-DP model shows an overall worse performance.

\textbf{Simulation 2}: When the model is misspecified, the S-GP-DP and S-GP-N models are again seen to yield an overall superior estimation performance, compared to the S-BP-DP and S-BP-N. The GP models deliver satisfactory estimation performances in terms of MSE, RB and MISE. The GP models attain the desired $95\%$ coverage across all sample sizes. Though S-BP-N attains the nominal CP, the coverage of S-BP-DP falls sharply with increasing sample size ($1\%$ for $n = 500$). The regression parameter estimates from the S-BP-N appear to be robust for all sample sizes, however, the estimation of the link function worsens acutely with increase in the sample size. Estimation of the link function and the regression parameters are observed to be poor under the S-BP-DP model. To summarize, estimates from the GP models are found to be robust under model misspecification. While the estimation of the link function under the S-BP-N model suffers for large sample sizes, the S-BP-DP model delivers poor performances across all sample sizes.

The poor performance of the BP-DP model can be attributed to the global nature of the BP basis, where the identifiability constraint ($g(-1) = 0$) impacts the overall flexibility of the link function estimation, unlike the local basis used in the GP method. Additionally, the BP-DP model faces the challenge of simultaneously estimating both the link function and the error density, whereas the BP-N model benefits from a fixed Gaussian assumption for the error distribution, enabling better recovery of the link function.

\section{Conclusions}
\label{s:conc}

Motivated by a cross-sectional PD study, we introduce a novel framework for estimating a clinically interpretable tooth-level disease risk index within a multistate model, accounting for spatially-referenced current status data. Our approach leverages advanced posterior sampling techniques, such as elliptical slice sampling and circulant embedding, to address computational challenges. Applied to the GAAD data, our model confirms higher PD risk among males and smokers \citep{genco2013risk}, but, consistent with our preliminary results, does not reveal a significant association between HbA1c levels and PD progression.

We estimate transition probabilities under the assumption that future transitions depend only on the current state. While incorporating sojourn times (\ie the duration a tooth remains in a particular state before transitioning) could be considered, this is challenging under the current status framework. Our model adopts a forward tracking structure, reflecting the progressive nature of PD \citep{coventry2000periodontal}, but could be extended to allow reversible transitions, as relevant in other diseases. We also assume independent censoring; relaxing this to allow for informative inspection times, ({\em e.g.}, via copula models \citep{zhao2015regression}), is complex and presents a separate scope of future study.
Recent advances in single index models using neural networks \citep{bietti2022SIM} do not enforce monotonicity of the link function, which is key for clinical interpretability. Integrating machine learning tools to estimate monotone SIMs is a promising future direction. Moreover, while advanced areal spatial models, such as the Directed Acyclic Graph Auto-Regressive model \citep{datta2019spatial} exist, we opt for a simpler inverse-Wishart approach to avoid complicating our existing framework on monotone link function and error density estimation. This aims to mitigate potential mixing issues that may arise from introducing additional complexities to the model.

Finally, our analysis of the GAAD dataset reveals its limitations, including only four subject-level covariates and a relatively small sample size, in light of recently available databases of larger dimensions. Extending our approach to larger cross-sectional studies like NHANES \citep{cdc1996plan} will require significant computational enhancements, but the availability of abundant covariates in such databases could facilitate techniques such as variable selection within the single index framework to construct an interpretable index.

\section*{Acknowledgements}
The authors acknowledge the Center for Oral Health Research at the Medical University of South Carolina for providing the motivating GAAD dataset, and partial funding support from NIH/NIDCR grants R21DE031879, and R01DE031134. 

\section*{Data Availability}
The motivating GAAD data that support the findings of this study are available at the GitHub repository, \href{https://github.com/das-snigdha/BayesSPMSM}{das-snigdha/BayesSPMSM}. Computing code is also available on this repository.

\bibliographystyle{apalike}
\bibliography{ref}

\newpage
\section*{\LARGE Supplementary Materials}
This supplement contains a preliminary analysis of the GAAD data, proof of results on model identifiability, blocked full conditional posterior distribution of parameters, details on estimation of state occupation and transition probabilities, additional details on the GAAD data analysis, and additional simulation results. Code for implementing our model is available at the GitHub repository, \href{https://github.com/das-snigdha/BayesSPMSM}{das-snigdha/BayesSPMSM}.

\appendix
\numberwithin{figure}{section}

\section{GAAD data: some specifics and preliminary analysis}
\label{s:data_prelim}

For an initial preliminary analysis, we converted the multistate responses of the GAAD data (see Section \ref{s:data_analysis} of the main document for details) into interval-censored, corresponding to tooth occupying state $3$, \ie\ tooth loss.

\begin{figure}[htp]
    \centering
    \includegraphics[width=\linewidth]{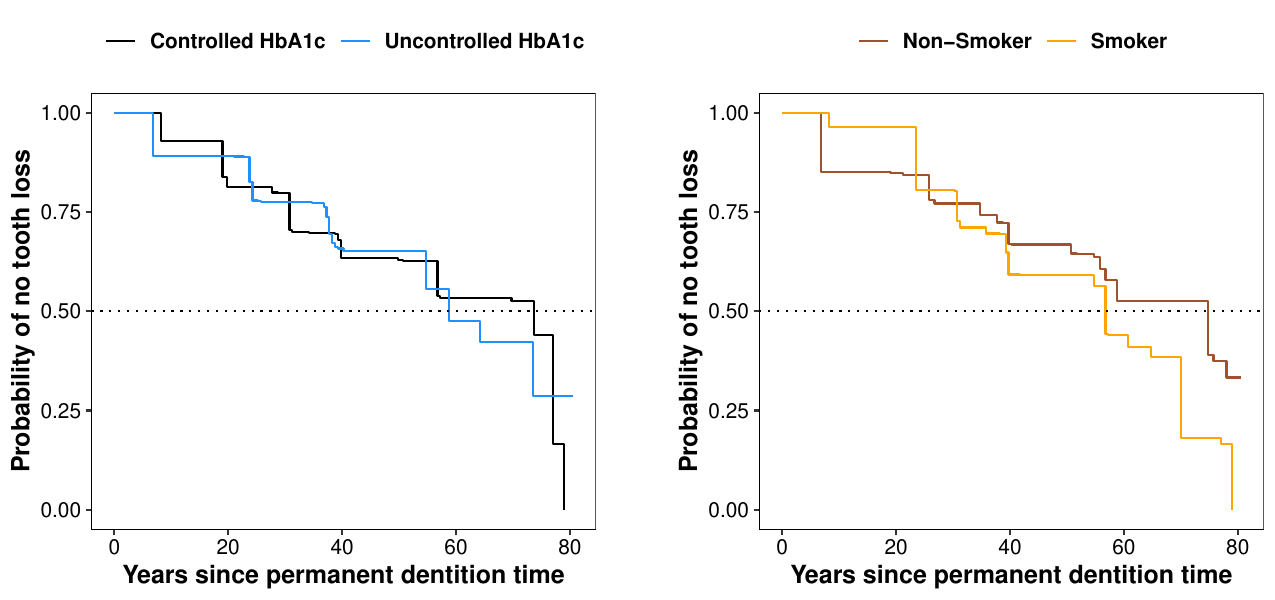}
    \caption{Nonparametric Turnbull estimates of tooth-level survival curves with tooth loss as the event of interest.}
    \label{fig:NP_MLE}
\end{figure}

Figure \ref{fig:NP_MLE} presents the nonparametric Turnbull estimates \citep{turnbull1976empirical}  of tooth-level survival curves, grouped by smoking status and HbA1c, revealing worse survival for smokers, but overlapping curves for the HbA1c levels.
Furthermore, a fitted accelerated failure time (AFT) model (with Weibull errors) incorporating squared BMI and interactions between smoking, gender and HbA1c along with the main covariates, reveals smoking, jaw indicator and smoking$\times$gender interaction to be strongly significant predictors. Additionally, squared BMI and smoking$\times$HbA1c interaction were marginally significant. This provides an indication for the presence of non-linearity in covariates. The absence of a clear unified strategy that accommodates non-linearity and interactions to yield a concise and interpretable single-number summary motivates us to develop a holistic AFT regression approach, incorporating monotone single index model within a progressive multistate modeling framework that facilitates efficient risk quantification on the entire tooth decay process via automatic consideration of non-linear main effects and higher order interactions through a scalar single index.

\section{Identifiability of parameters}
\label{s:proofs}
In the following, we provide the proofs of Propositions \ref{p:variance_random_effects} and \ref{p:identifiability}, which establishes the identifiability of the variance components and fixed effects parameters of our proposed model.

\subsection{Proof of Proposition 1}
\label{s:proof1}
We start by recalling from Section \ref{s:identifiability} of the main document, that our AFT model can be concisely written as
\begin{equation*}
    \uy_i = \uZ \ub_i + \ug_i + \uepsilon_i, \quad\quad i = 1,2, \ldots, n
\end{equation*}
where $\uZ = [\mathbb{I}_{m/2}\ \Tilde{\mathbb{I}}_{m/2}]^\top$, and $\mathbb{I}_{m/2}$, $\Tilde{\mathbb{I}}_{m/2}$ denote $m/2 \times m/2$ identity matrix and anti-diagonal matrix with unit entries, respectively. 
The covariance matrix of $\uy_i$ is then given by $\Sigma_y = \uZ \Sigma_b \uZ^\top + \Sigma_\epsilon$, where $\Sigma_\epsilon = \sigma_\epsilon^2 \mathbb{I}_m$, $\sigma_\epsilon^2 > 0 $ denotes the common variance of the errors $\epsilon_{ij}$ and $\Sigma_b$ denotes the covariance matrix of the random effects $\ub_i$. 
Further, let $(\Sigma_b)_{jj^\prime} = \sigma_{b,j}^2 \mathds{1}_{\{j = j^\prime\}} + \sigma_{b,jj^\prime} \mathds{1}_{\{j \neq  j^\prime\}}$, where $\sigma_{b, j}^2 > 0$ for all $j$ and $\sigma_{b, jj^\prime} = \sigma_{b, j^\prime j} \in \mathbb{R}$ for all $j \neq j^\prime$.

Suppose $\Sigma(\utheta)$ denotes a matrix parameterized by a vector $\utheta$ in the space $\Theta = \{\utheta : \Sigma(\utheta) \text{ is symmetric and positive definite}\}$. Let $\Sigma_\epsilon$ and $\Sigma_b$  be parameterized by $\utheta_\epsilon = \{\sigma_\epsilon^2\} \in \Theta_\epsilon$  and $\utheta_b = \{ \sigma_{b, j}^2, \sigma_{b, jj^\prime} : 1\leq j < j^\prime \leq m/2 \} \in \Theta_b$, respectively. Then, it suffices to prove that
$\Sigma_y(\utheta_b, \utheta_\epsilon) = \Sigma_y(\utheta^\prime_b, \utheta^\prime_\epsilon)$ implies $\utheta_b = \utheta^\prime_b$ and $\utheta_\epsilon = \utheta^\prime_\epsilon$ for all $\utheta_b, \utheta^\prime_b \in \Theta_b $ and $\utheta_\epsilon, \utheta^\prime_\epsilon \in \Theta_\epsilon$.
To that end, note that
\begin{equation}
\label{eq:var_y}
\text{Var}(y_{ij}) = (\Sigma_y)_{jj}=
\begin{cases}
	\sigma_{b, j}^2 + \sigma_\epsilon^2, & \text{if } 1\leq j \leq \frac{m}{2}\\
    \sigma_{b, m-j+1}^2 + \sigma_\epsilon^2, & \text{if } \frac{m}{2} < j \leq m
\end{cases}
\end{equation}
and for $j < j^\prime$, we have
\begin{equation}
\label{eq:cov_y}
\text{Cov}(y_{ij}, y_{ij^\prime}) = (\Sigma_y)_{j j^\prime}=
\begin{cases}
	\sigma_{b, jj^\prime}, & \text{if } 1\leq j < j^\prime \leq \frac{m}{2}; \ j^\prime \neq m - j +1\\
    \sigma_{b, (m-j+1)(m-j^\prime+1)}, & \text{if } \frac{m}{2} < j < j^\prime \leq m; \ j^\prime \neq m - j +1\\
    \sigma_{b, j(m-j^\prime+1)}, & \text{if } 1 \leq j \leq \frac{m}{2} < j^\prime \leq m \\
    \sigma_{b, j}^2, & \text{if } j^\prime = m - j +1
\end{cases} 
\end{equation}
Then, $\Sigma_y(\utheta_b, \utheta_\epsilon) = \Sigma_y(\utheta^\prime_b, \utheta^\prime_\epsilon)$ implies $\utheta_b = \utheta^\prime_b$ using \eqref{eq:cov_y} and $\utheta_\epsilon = \utheta^\prime_\epsilon$ using \eqref{eq:var_y}.

\subsection{Proof of Proposition 2}
\label{s:proof2}

We begin by noting that our observed current status data is represented as $\{ \ux_{ij}, C_{ij}, S_{ij}(C_{ij}) \}$, where $\ux_{ij}$ denotes the vector of covariates, $C_{ij}$ denotes the random inspection times and $S_{ij}(C_{ij})$ denotes the disease state occupation at the inspection time. Clearly, $S_{ij}(C_{ij})$ takes values in $\mathcal{K} = \{0,1, \ldots, K\}$ and let $\uc_i = (c_{i1}, \ldots, c_{im}) $ denote the observed values of the inspection times.
Under the assumption of independent censoring, the likelihood function for our model is given by
\begin{align*}
    L(\utheta, \ueta) 
    & = \prod_{i=1}^n p_{\uC}(c_{i1}, \ldots, c_{im}) \, p_{\uS}\left(s_{i1}(c_{i1}), s_{i2}(c_{i2}), \ldots, s_{im}(c_{im})\mid \utheta, \ualpha, \upsi\right)\, \\
    & \propto \prod_{i=1}^n  p_{\uS}\left(\us_{i}\mid \utheta, \ualpha, \upsi\right)\,,
\end{align*}
where $p_{\uC}$ denotes the joint distribution of the random inspection times $\uC_i = (C_{i1}, \ldots, C_{im})$ and $p_{\uS}(\cdot \mid \utheta, \ualpha, \upsi)$ denotes the joint distribution of state occupation given the observed inspection times (\ie \ $\uC_i = \uc_i$), $\uS_i = \{S_{ij}(c_{ij}): j =1, 2,\ldots, m \}$, $i = 1,2, \ldots, n$ by marginalizing over all latent variables. $p_{\uS}(\cdot \mid \utheta, \ualpha, \upsi)$ is parameterized by the fixed effects $\utheta = (g, \ubeta)$, the parameters $\ualpha$ of the Dirichlet distribution on the relative disease increment times $\uR_{ij}$, and $\upsi = (\Sigma_b, F)$, where $\Sigma_b$ and $F$ denote the fixed covariance matrix of the random effects $\ub_i$ and baseline distribution of the errors, respectively. 
Note that $\utheta \in \Theta =\{(g, \ubeta): g \in \mathcal{C}_\mathcal{M}, \ubeta \in \mathcal{S}_{p-1}\}$ and $\ualpha \in (0, \infty)^K$.
Under independent censoring, $p_{\uC}$ is free of the parameters involved in the multistate model and hence the censoring distribution does not contribute to the likelihood function of our model parameters.

Let $\ualpha_1, \ualpha_2 \in (0, \infty)^K$ and $\utheta_1, \utheta_2 \in \Theta\,$ such that 
\begin{equation}
\label{eq:joint_S}
    p_{\uS}(\us \mid \utheta_1, \ualpha_1, \upsi) = p_{\uS}(\us \mid \utheta_2, \ualpha_2, \upsi)\,.
\end{equation}
To demonstrate identifiability of the fixed effects parameters, it suffices to show that \eqref{eq:joint_S} implies $\utheta_1 = \utheta_2$ for all $\utheta_1, \utheta_2 \in \Theta$.
Note that \eqref{eq:joint_S} implies
\begin{equation}
\label{eq:marginal_S}
    p_{S_j}(s_j \mid \utheta_1, \ualpha_1, \upsi) = p_{S_j}(s_j \mid \utheta_2, \ualpha_2, \upsi), \quad \text{for all }j = 1, 2, \ldots, m,
\end{equation}
where, $p_{S_j}(\cdot \mid \utheta, \ualpha, \upsi)$ denotes the marginal distribution of $S_{ij}(c_{ij})$, the disease state occupied by tooth $j$ at its observed inspection time $c_{ij}$, for each subject $i = 1, 2, \ldots n$. Thus, it suffices to work with the disease state of one tooth, say $S_{i1}$ at its inspection time $c_{i1}$ for any subject $i$. Let $\sigma_{b,1}^2 = (\Sigma_b)_{11}$ denote the marginal variance of the random effect for this tooth. Henceforth, we shall drop the subscripts $(i,1)$ from the random variables for clarity of notation.

From equation (\ref{eq:observed_states}) of the main document:
\begin{equation}
\label{eq:observed_states_s}
S(c) = k \quad \textrm{if}\quad  {T} \sum_{l=0}^k  {R^{\,(l)}} \leq   {c  } <  {T } \sum_{l=0}^{k+1}  {R ^{\,(l)}}, \quad k = 0,1,\ldots, K\,,
\end{equation} 
where $R ^{\,(0)} := 0$ and $R ^{\,(K+1)} := \infty$.
From \eqref{eq:observed_states_s}, we observe that the marginal distribution of $S(c)$ is a categorical distribution on $\mathcal{K} = \{0,1,\ldots, K\}$ with probabilities $\{p_k(\utheta, \ualpha, \upsi) = P_{\utheta, \ualpha, \upsi} \left(S(c) = k  \right): k \in \mathcal{K}\}$. \eqref{eq:marginal_S} implies 
\begin{equation}
\label{eq:cat_S}
p_k(\utheta_1, \ualpha_1, \upsi) = p_k(\utheta_2, \ualpha_2, \upsi), \quad \text{for all } k \in \mathcal{K}\,.
\end{equation}
Observe that for $k = K$, we have $S (c ) = K$ if ${T } \leq   {c  }$,
since $\uR $ follows a Dirichlet distribution and $\sum_{l=0}^K  {R ^{\,(l)}} = 1 $ with probability 1. 
Therefore, 
\begin{align}
    \label{eq:prob_K}
    p_K(\utheta, \ualpha, \upsi) 
    & =  P_{\utheta, \ualpha, \upsi}\left(T  \leq c  \right) = F \left(\log c  - b  - g(\ux ^\top\ubeta) \right)
\end{align}
\noindent Letting $\mathbb{S}_{K-1}$ denote a $K$-dimensional simplex and writing $p_K(\utheta)$ explicitly, we have
\begin{align}
    \nonumber
    p_K(\utheta, \ualpha, \upsi) 
    & = P\left(S(c) = K \mid \utheta, \ualpha, \upsi \right) \\
    \nonumber
    & = \int_\mathbb{R} \int_{\mathbb{S}_{K-1}} P_{\utheta, \ualpha, \upsi}\left(S (c ) = K \mid \ub, \uR\right)  \frac{1}{\sigma_b}\phi \left(\frac{b}{\sigma_b} \right) \, f_{\uR}\left( \uR \mid \ualpha \right) \, d\uR \  db \\[2ex]
    \nonumber
    & = \int_\mathbb{R}  F \left(\log c - b - g(\ux^\top\ubeta) \right)   \frac{1}{\sigma_b}\phi \left(\frac{b}{\sigma_b} \right) \left\{ \int_{\mathbb{S}_{K-1}} f_{\uR}\left( \uR \mid \ualpha \right)\, d\uR \right\} \,  db \\[2ex]
    \label{eq:p_K}
    & = \int_\mathbb{R} F \left(\log c - b - g(\ux^\top\ubeta) \right) \frac{1}{\sigma_b}\, \phi \left(\frac{b}{\sigma_b} \right) \ db \,,
\end{align}
where $\phi(\cdot)$ and $f_{\uR}(\cdot \mid \ualpha)$ denote the probability density functions of a standard Gaussian distribution and a Dirichlet distribution with parameters $\ualpha$, respectively.

Next, define a function $$h(z) = \int_\mathbb{R} F \left(\log c - b - z \right) \frac{1}{\sigma_{b}}\, \phi \left(\frac{b_1}{\sigma_{b}} \right) \ db_1\,.$$ We claim that $h(z)$ is a strictly decreasing function of $z$. As a proof, we first note that $h(z)$ is non-increasing in $z$, since $F(\log c - b - z)$ is a non-increasing function of $z$, for all $b\in \mathbb{R}$. The proof follows by contradiction. Let, if possible, $h(z_1) = h(z_2)$ for $z_1 < z_2$.
Let $$\psi(y) = \frac{1}{\sigma_b} \left[\phi \left( \frac{y+z_1 - \log c}{\sigma_{b}} \right) - \phi \left( \frac{y+z_2 - \log c}{\sigma_{b}} \right) \right], \quad y\in \bbR,$$
and $z^* = \log c - (z_1 + z_2)/2$, and note that
\begin{align}
    \psi(z^* - y) &= -\, \psi(z^* + y), \quad y \in \mathbb{R} \label{eq:psi_1}\\
    \psi(z^* + y) &>0, \quad y>0 \label{eq:psi_2}
\end{align}
Then,
\begin{equation}
\begin{split}
\label{eq:hx1_hx2}
    0 &= h(z_1) - h(z_2)\\
    & = \int_\mathbb{R} F \left(\log c - b_1 - z_1 \right) \frac{1}{\sigma_{b}} \phi \left(\frac{b_1}{\sigma_{b}} \right) \ db_1 - \int_\mathbb{R} F \left(\log c - b_1 - z_2 \right) \frac{1}{\sigma_{b}} \phi \left(\frac{b_1}{\sigma_{b}} \right) \ db_1\\
    & = \int_\mathbb{R} F \left(y \right) \frac{1}{\sigma_b} \left[ \phi \left(\frac{y + z_1 - \log c}{\sigma_{b}} \right) -  \phi \left(\frac{y + z_2 - \log c}{\sigma_{b}} \right) \right] dy \\
    & = \int_\mathbb{R} F (y) \ \psi(y) \ dy \\
    & = \int_{-\infty}^{z^*} F (y )\ \psi(y) \ dy \ + \ \int_{z^*}^\infty F (y )\ \psi(y) \ dy \\
    & = \int_{-\infty}^{z^*} \{ F (y) - F(2z^* - y)\} \,\psi(y) \, dy,  \quad \qquad \text{since } \psi(2z^* - y) = - \psi(y)\\
    & = \int_{0}^\infty \{ F (z^* + y) - F(z^* - y)\}  \ \psi(z^* + y) \ dy, \qquad \quad \text{using } \eqref{eq:psi_1} \,.
\end{split}
\end{equation}
Using \eqref{eq:psi_2}, \eqref{eq:hx1_hx2} implies $F (z^* + y) = F(z^* - y)$, for all $y >0$, which shows that $F$ is symmetric about $z^*$ and contradicts the fact that $F$ is a CDF. 

Finally, \eqref{eq:cat_S} implies $p_K(\utheta_1, \ualpha_1, \upsi) = p_K(\utheta_2, \ualpha_2, \upsi)$, which is equivalent to $ h(g_1(\ux^\top \ubeta_1)) = h(g_2(\ux^\top \ubeta_2))\,$.
By invertibility of strictly monotone functions, we have $g_1(\ux^\top \ubeta_1) = g_2(\ux^\top \ubeta_2)$, which further implies $g_1 = g_2$ and $\ubeta_1 = \ubeta_2$, by Theorem 1 from \cite{linkulasekera}.

\section{Posterior Computation}
\label{s:posteriors}
This section enlists detailed derivations of the blocked full conditional distributions of the parameters and the latent variables of our model, along with their posterior sampling schemes. Recall from Section \ref{s:prior_errors} of the main document that a Dirichlet process (DP) mixture prior of Gaussians on the baseline density $f$ of the errors posits that 
\begin{align}
\label{eq:DPM}
    \epsilon_{ij} \mid \mu_{ij}, \sigma^2_{ij} &\sim \mathcal{N}(\mu_{ij}, \sigma^2_{ij}), & (\mu_{ij}, \sigma^2_{ij}) \mid G &\sim G, & G &\sim DP(\gamma, G_0).
\end{align}
Using the atoms $(\mu_{ij}, \sigma^2_{ij})$ of the DP mixture prior of Gaussians on the density of the random errors, we can express our AFT model given in equation (\ref{eq:AFT}) of the main document as, 
$$\log T_{ij} = g(\ux_{ij}^\top \ubeta) + b_{ij} + \mu_{ij} + \sigma_{ij}\Tilde{\epsilon}_{ij}\ , \quad \Tilde{\epsilon}_{ij} \sim \mathcal{N}(0,1),$$ 
where $g(\ux_{ij}^\top \ubeta) \approx \sum_{l=0}^L \xi_l \psi_l(\ux_{ij}^\top \ubeta)$ using equation (\ref{eq:basis_expansion}) of the main document.
The subscripts $i$ and $j$ denote the subjects and teeth of each subject respectively in all successive subsections, $i = 1, 2, \ldots, n$, $j = 1, 2, \ldots, m$.

\subsection{Regression parameters}
\label{s:post_beta}
Define 
\begin{equation}
\label{eq:mean_post_beta}
    u_{ij} = \log T_{ij} - b_{ij} - \mu_{ij}, \quad \text{for each } i, j.
\end{equation}
The conditional posterior of $\widetilde \ubeta$ is given by,
$p \big(\,\widetilde \ubeta \mid \ldots \, \big) \propto \  \mathcal{N}\big(\widetilde \ubeta \,; 0, \sigma_\beta^2 \big) \ L_{\beta} \big(\,\widetilde \ubeta \, \big),$
where, 
$$\log L_{\beta} \big(\,\widetilde \ubeta \, \big) = - \frac{1}{2}\sum_{i,j} \frac{1}{\sigma_{ij}^2} \left(u_{ij} - \sum_{l=0}^L \xi_l \,\psi_l \big(\ux_{ij} ^\top \frac{\widetilde \ubeta}{\|\widetilde \ubeta \|}\big) \right)^2 .$$
$\widetilde \ubeta$ is sampled using the elliptical slice sampler \citep[ESS,][]{murray2010elliptical}, particularly suited for posterior sampling in models with multivariate Gaussian priors.

\subsection{Monotone link function}
\label{s:post_link}
Let $u_{ij}$ be defined as \eqref{eq:mean_post_beta}. Define $\Psi$ be an $N \times (L+1)$ matrix with elements $\Psi_{r l} = \psi_l(\ubeta^T \ux^{*}_r) / \sigma_{r}^{*}$, where $N = mn$, and

\vspace{-1cm}

\begin{align*}
    (\ux^{*}_{1}, \ldots, \ux^{*}_{N}) & = (\ux_{11}, \ux_{12},\ldots, \ux_{1m}, \ldots, \ux_{n1}, \ux_{n2}, \ldots, \ux_{nm})\\
    (\sigma^{*}_{1}, \ldots, \sigma^{*}_{N}) & = (\sigma_{11}, \sigma_{12}, \ldots, \sigma_{1m}, \ldots, \sigma_{n1}, \sigma_{n2}, \ldots, \sigma_{nm})
\end{align*}
are simply the stacked covariates and error standard deviations over $i$ and $j$. Similarly stack $\uu = (u_{11}, u_{12}, \ldots, u_{1m}, \ldots, u_{n1}, u_{n2}, \ldots, u_{nm})$.

The coefficients $\uxi$ in the basis expansion of the monotone link function given in equation (\ref{eq:basis_expansion}) of the main document, have a zero-mean Gaussian prior with covariance matrix $\tau^2 K$, where $K_{ij} = k(u_i - u_j)$, $k(\cdot)$ is the stationary Matérn covariance kernel with smoothness parameter $\nu > 0$ and length-scale parameter $\ell > 0$.  The most general definition of the Matérn covariance kernel is the following:
$$k(r) = \frac{2^{1- \nu}}{\Gamma(\nu)} \left( \frac{\sqrt{2\nu} \, r }{\ell} \right)^{\nu} K_{\nu} \left( \frac{\sqrt{2\nu} \, r }{\ell} \right),$$
where $K_{\nu}$ is the modified Bessel function of the second kind. The parameter $\tau$ controls the prior signal to noise ratio. A non-informative inverse-gamma $\mathcal{IG}(a_\xi, b_\xi)$ prior is placed on $\tau^2$. 
The conditional posterior of the parameters involving estimation of the monotone link function is given by,
\begin{align*}
    p(\bm \xi, \tau \mid \ldots) \propto (\tau^2)^{-(a_\xi+1)} e^{-{b_\xi}/\tau^2 } (\tau^2)^{-(L+1)/2} e^{\,-{\bm \xi}^\top K^{-1}{\bm \xi} \,/\, (2\tau^2)} \ \mathbb{J}_\eta(\bm \xi) \ e^ {\,- \|\uu - \Psi{\bm \xi}\|^2 /2},
\end{align*}
where $\mathbb{J}_\eta(\uxi) = \prod_{l=0}^{L} (1+e^{\, - \eta\,\xi_l})^{-1}$.
The sampling algorithm proceeds as follows:
\begin{enumerate}
    \item Update $\uxi$:
    The full conditional of $\uxi$ is given by
    $$p (\uxi \mid \ldots \, ) \propto \  \mathcal{N}(\uxi \,;0, \tau^2 K) \ L_{\xi} (\uxi),$$
    where $L_{\xi} (\uxi) = e^ {\,- \|\uu - \Psi{\bm \xi}\|^2 /2}\ \mathbb{J}_\eta(\uxi)$.
    Posterior samples of $\uxi$ are obtained using the ESS scheme by \cite{murray2010elliptical} by sampling a proposal from its prior $\mathcal{N}(0, \tau^2 K)$ using the algorithm by \cite{wood1994simulation}.

    \item Update $\tau^2$: The full conditional of $\tau^2$ is $\mathcal{IG} \big(a_\xi+(L+1)/2,\ b_\xi + {\bm \xi}^\top K^{-1}{\bm \xi}/2 \big)$.
\end{enumerate}


\subsection{Random effects}
With slight abuse of notation, let $u_{ij} = \log T_{ij} - g(\ux_{ij}^\top \ubeta) - \mu_{ij}$ for each $i$, $j$. For each $i$, let 
\begin{align*}
    \uu_i^{(1)} &= (u_{i1}, u_{i2}, \ldots, u_{im/2})^\top & \Omega_{\epsilon, i}^{(1)} & = \text{diag} (1/\sigma^2_{i1}, 1/\sigma^2_{i2}, \ldots, 1/\sigma^2_{im/2})\\
    \uu_i^{(2)} &= (u_{im}, u_{i(m-1)},\ldots, u_{i(m/2+1)})^\top &
    \Omega_{\epsilon, i}^{(2)} & = \text{diag} (1/\sigma^2_{im}, 1/\sigma^2_{i(m-1)}, \ldots, 1/\sigma^2_{i(m/2+1)})
\end{align*}
The conditional posterior of the random effects vector $\ub_i = (b_{i1}, \ldots , b_{im/2})^\top$ for each $i$, is Gaussian with mean vector $\bm Q_b^{-1}\ua_b$ and covariance matrix $\bm Q_b^{-1}$, where 
\begin{align*}
    \bm Q_b & = \Sigma_b^{-1} + \Omega_{\epsilon, i}^{(1)} + \Omega_{\epsilon, i}^{(2)}, &
    \ua_b & = \Omega_{\epsilon, i}^{(1)} \uu_i^{(1)} + \Omega_{\epsilon, i}^{(2)} \uu_i^{(2)}.
\end{align*}

\subsection{Covariance matrix of random effects}
The conditional posterior of $\Sigma_b$ is an inverse-Wishart distribution, $\mathcal{IW} \left(c+n, S + \sum_{i=1}^n \ub_i \ub_i^\top \right)$, where $c = m/2 +2$ and $S = (E_W - \rho W)^{-1}$, as defined in Section \ref{s:prior_spatial_random_effects} of the main document.

\subsection{Density of errors}
\label{s:density_post}

The stick-breaking construction \citep{sethuraman1994constructive} allows us to express $G$ in \eqref{eq:DPM} as 
\begin{align*}
    G(\mu, \sigma^2) & = \sum_{h=1}^\infty \pi_h \delta_{(\phi_h, \, s^2_h)}(\mu, \sigma^2), & 
    \left(\pi_h\right)_{h=1}^\infty &\sim \text{GEM}(\gamma), & \left(\phi_h, s^2_h \right)_{h=1}^\infty &\sim G_0, 
\end{align*}
where GEM stands for Griffiths, Engen and McCloskey \citep{pitman2002poissondirichlet}, and $\delta_{(\phi,\, s^2)}(\cdot)$ is the Dirac measure at $(\phi, s^2)$. Here, $(\phi_h, s_h^2)_{h=1}^\infty$ denote the distinct values of the atoms $(\mu_{ij}, \sigma_{ij})$ used in \eqref{eq:DPM}.
Posterior sampling for the parameters of the DP mixture prior on the baseline density $f$ is performed using a blocked Gibbs algorithm by \cite{ishwaran2001gibbs} that uses a finite truncation for the number of
mixture components upto a large number H.

Recall that $\epsilon_{ij} = \log T_{ij} - b_{ij} - g(\ux_{ij}^\top\ubeta)$, for each $i$, $j$. The hierarchical representation of a DP mixture prior of Gaussians on $\epsilon_{ij}$ with a normal-inverse gamma base distribution $G_0$ and a gamma prior on the concentration parameter $\gamma$ is as follows:
\begin{align*}
    \epsilon_{ij} \mid z_{ij}, \uphi, \us & \sim \mathcal{N}(\phi_{z_{ij}}, s_{z_{ij}}^2), \quad \text{for each } i, j.\\
    z_{ij} \mid \upi & \sim \text{Cat}(\pi_1, \pi_2, \ldots, \pi_H)\\
    \upi \mid \gamma & \sim \text{GEM} (\gamma) \\
    \phi_h, s_h^2 & \sim \mathcal{NIG}(\mu_\epsilon, \nu_\epsilon, a_\epsilon, \lambda_\epsilon), \quad h = 1, 2, \ldots, H.\\
    \gamma & \sim \mathcal{G}(a_\gamma, b_\gamma)
\end{align*}
where $\uphi = (\phi_1, \phi_2, \ldots, \phi_H)$, $\us = (s^2_1, s^2_2, \ldots, s^2_H)$ and $\upi = (\pi_1, \pi_2, \ldots, \pi_H)$.
Note that the prior on $\upi$ can be explicitly written as $\pi_1 = \pi_1^\prime$, $\pi_h = \pi^\prime_h \prod_{l<h}(1-\pi^\prime_l)$ for $1<h<H$, and 
$\pi_H = \prod_{h=1}^{H-1} (1-\pi_h)$, where 
$\pi_l^\prime \mid \gamma \sim \text{Beta}(1, \gamma)$, $l=1, \ldots, H-1$. The atoms $(\mu_{ij}, \sigma_{ij}^2)$ can be obtained using the distinct atoms $(\phi_h, s_h^2)_{h=1}^H$ and the cluster labels $z_{ij}$ as follows, $$\mu_{ij} = \phi_{z_{ij}}, \quad {and} \quad \sigma_{ij}^2 = s_{z_{ij}}^2, \quad \text{for each } i,j.$$

Blocked posterior updates of the parameters are as follows:
\begin{enumerate}
    \item Update $\uz$. For each $i$, $j$,
    $$z_{ij} \mid \ldots \sim \text{Cat}\,(\pi^*_1, \ldots, \pi^*_H),$$
    where $\pi^*_h = \pi_h \, \mathcal{N}(x_{ij}\,; \phi_h, s_h^2)\big/ \sum_{l=1}^H \pi_l \, \mathcal{N}(x_{ij}\,; \phi_l, s_l^2)$ for each $h$.

    \item Update $\upi$. For $h = 1, 2, \ldots, H-1$,
    $$\pi^\prime_h \mid \ldots \sim \text{Beta} \,\big(n_h+1, \gamma + \sum_{l>h} n_l \big),$$
    where $n_h = \sum_{i,j}\mathds{1}_{\{z_{ij} = h\}}$ for each $h$. Set $\pi_1 = \pi^\prime_1$, $\pi_h = \pi^\prime_h \prod_{l<h}(1-\pi^\prime_l)$ for $1<h<H$, and $\pi_H = \prod_{l=1}^{H-1} (1-\pi^\prime_l)$.

    \item Update $\uphi, \us$. For $h = 1, 2, \ldots, H$,
    \begin{align*}
        s_h^2 \mid \ldots &\sim \mathcal{IG}(\,\Tilde{a}_{\epsilon}, \Tilde\lambda_{\epsilon}\,) \\
        \phi_h \mid s_h^2, \ldots &\sim \mathcal{N}(\,\Tilde{\mu}_\epsilon,\ s_h^2 / \Tilde\nu_\epsilon\,)
    \end{align*}
    where 
    \begin{align*}
        \Tilde{a}_{\epsilon} & = a_\epsilon + n_h/2, &
        \Tilde{\lambda}_\epsilon & = \lambda_\epsilon \ + \ \frac{1}{2}\, \bigg[ \sum_{i, j : z_{ij} = h} (\epsilon_{ij} - \Bar{\epsilon}_h)^2 \ + \ \frac{\nu_\epsilon n_h}{\nu_\epsilon + n_h} (\Bar{\epsilon}_h - \mu_\epsilon)^2 \bigg],\\
        \Tilde{\nu}_\epsilon &= \nu_\epsilon + n_h, & 
        \Tilde{\mu}_\epsilon &= \frac{\nu_\epsilon \mu_\epsilon + n_h \Bar{\epsilon}_h}{\nu_\epsilon + n_h}, \quad \quad \Bar{\epsilon}_h = \frac{1}{n_h}\sum_{i, j : z_{ij} = h} \epsilon_{ij}.
    \end{align*}

    \item Update $\gamma$.
    $$\gamma \mid \ldots \sim \mathcal{G} \bigg(a_\gamma + H-1,\, b_\gamma - \sum_{h<H} \log (1-\pi^\prime_h) \bigg).$$
\end{enumerate}

\subsection{Time to a missing tooth}
\label{s:post_logT}
Let $s_{ij} = S_{ij}(c_{ij})$ denote the state of tooth $j$ of subject $i$ at its observed inspection time $C_{ij} = c_{ij}$.
The conditional posterior distribution of $\log T_{ij}$ for each $i$, $j$, is a truncated Gaussian with mean $g(\ux_{ij}^\top \ubeta) + b_{ij}+ \mu_{ij}$ and variance $\sigma_{ij}^2$, where the truncation region is given by 
$$\mathcal{C}_{\log T_{ij}}= \left\{\log T_{ij}   \in \bbR  : \ \frac{c_{ij}}{\sum_{l=1}^{s_{ij}+1} R_{ij}^{\,(l)}} \ < \ T_{ij} \ \leq \ \frac{c_{ij}}{\sum_{l=1}^{s_{ij}} R_{ij}^{\,(l)}} \right\}.$$

\subsection{Relative disease increment times}
The conditional posterior distribution of the relative time increment to each disease state $\uR_{ij} = \left(R_{ij}^{\,(1)}, R_{ij}^{\,(2)}, \ldots, R_{ij}^{\,(K)}\right)$ for each $i$, $j$, is a truncated Dirichlet distribution with parameters $\left(\alpha_1, \ldots, \alpha_K\right)$ and the region of truncation given by 
$$\mathcal{C}_{\uR_{ij}} = \left\{\uR_{ij} \in \mathbb{S}_{K-1}: \ \frac{c_{ij}}{\sum_{l=1}^{s_{ij}+1} R_{ij}^{\,(l)}} \ < \ T_{ij} \ \leq \ \frac{c_{ij}}{\sum_{l=1}^{s_{ij}} R_{ij}^{\,(l)}} \right\},$$
where $\mathbb{S}_{K-1}$ denotes a simplex of dimension $K$, and $s_{ij}$ is as defined in Section \ref{s:post_logT}. We define $V_{ij}^{(k)} = \sum_{l=1}^k R_{ij}^{(l)}$, $k = 1, 2, \ldots, K-1$, and obtain samples from the posterior distribution using properties of the Dirichlet distribution and depending on $s_{ij}$, as follows:

\begin{enumerate}
    \item $s_{ij} = 0$. 
    \begin{enumerate}
        \item[a.] Sample $R_{ij}^{(1)} \sim$ Beta$(\alpha_1, \alpha_2 + \cdots +\alpha_K)$ truncated on $(c_{ij} / T_{ij}, 1)$.
        \item[b.] Sample  $(R_{ij}^{(2)}, \ldots, R_{ij}^{(K)} ) \sim (1-R_{ij}^{(1)}) \text{ Dir}(\alpha_2, \ldots, \alpha_K)$.
    \end{enumerate}
    
    

    \item $s_{ij} = k$, $k = 1, 2, \ldots, K-2$.
    \begin{enumerate}
        \item[a.]  Sample $V_{ij}^{(k)} \sim$ Beta$\big(\sum_{l=1}^k\alpha_l, \, \sum_{l=k+1}^K\alpha_l \big)$ truncated on $(0, c_{ij}/T_{ij})$. 
        \item[b.]  Sample $ U_{ij} \sim  \text{ Beta}\big(\alpha_{k+1},\, \sum_{l=k+2}^K\alpha_l \big)$ truncated on $\big(\, {(c_{ij}/T_{ij} - V_{ij}^{(k)})}/{(1-V_{ij}^{(k)})},1 \big)$. 
        
        \item[c.] Set $R_{ij}^{(k+1)}  = U_{ij}(1-V_{ij}^{(k)})$.
        \item[d.] Sample $(R_{ij}^{(1)} , \ldots , R_{ij}^{(k)}) \sim V_{ij}^{(k)} \ {\rm Dir(\alpha_1, \ldots, \alpha_k)}$.
        \item[e.] Sample $(R_{ij}^{(k+2)} , \ldots , R_{ij}^{(K)}) \sim (1-V_{ij}^{(k)}-R_{ij}^{(k+1)}) \ {\rm Dir(\alpha_{k+2}, \ldots, \ \alpha_K)}$.

    \end{enumerate}
    
    \item $s_{ij} = K-1$.

    \begin{enumerate}
        \item[a.] Sample $V_{ij}^{(K-1)}\sim$ Beta$(\alpha_1 + \cdots + \alpha_{K-1}, \, \alpha_K)$ truncated on $(0, c_{ij}/T_{ij})$
        \item[b.] Set $R_{ij}^{(K)}=1-V_{ij}^{(K-1)}$.
        \item[c.] Sample $(R_{ij}^{(1)} , \ldots , R_{ij}^{(K-1)}) \sim V_{ij}^{(K-1)} \ {\rm Dir(\alpha_1, \ldots, \alpha_{K-1})}$.
    \end{enumerate}


    \item $s_{ij} = K$.

    \begin{enumerate}
        \item[a.] Sample $(R_{ij}^{(1)} , \ldots , R_{ij}^{(K)})\sim {\rm Dir}(\alpha_1, \ldots, \alpha_K)$.
    \end{enumerate}

\end{enumerate}

\subsection{Dirichlet parameters of the relative disease increment times}
The conditional posterior of parameters $\ualpha = (\alpha_1, \ldots, \alpha_K)$ of the Dirichlet distribution on $\uR_{ij}$ is given by 
$$p(\ualpha \mid \ldots) \propto \prod_{k=1}^K \mathcal{G}(\alpha_k \,; a_\alpha, b_\alpha) \ \prod_{i,j}\text{Dir} (\uR_{ij}\,; \ualpha)$$ 
Samples from this non-standard posterior are obtained by using a random walk Metropolis within Gibbs with a Gaussian proposal whose variance is tuned to have a acceptance ratio between $0.2-0.4$.

\section{Estimation of state occupation and transition probabilities}
\label{s:SOP_TP_supp}

The marginal distribution of the time to a missing tooth $T_{ij}$, for each $i$, $j$ as described in Section \ref{s:probabilities} of the main document, is given by
\begin{equation}
\label{eq:dist_T_supp}
    \log T_{ij} \sim \sum_{h=1}^\infty \pi_h \, \mathcal{N}\left(\, g(\ux_{ij}^\top \,\ubeta) + \phi_h, \ \sigma^2_{b,j} + s^2_h \,\right),
\end{equation}
where $\sigma^2_{b,j} = \uz_j^\top \Sigma_b\, \uz_j$ and  $\uz_j^\top$ denotes the $j$th row of $\uZ$.
The number of mixture components in \eqref{eq:dist_T_supp} is truncated upto a large number $H$, as described in Section \ref{s:density_post}.

Let $\{\ubeta^{(d)}, \uxi^{(d)}, \Sigma_b^{(d)}, \upi^{(d)},\uphi^{(d)}, \us^{(d)}, \ualpha^{(d)} : d = 1, 2, \ldots M\}$ denote the set of $M$ posterior samples of the model parameters. In the following, we outline the steps for the estimation of the state occupation probability (SOP) and transition probability (TP) for tooth $j$ of a subject with any given choice of covariates, corresponding to each posterior sample $d$, $d = 1,2,\ldots,M$. Henceforth, the subscripts $ij$ are dropped for clarity of notation.

\vspace{0.5cm}

\noindent \textbf{Step 1.} Construct $g^{(d)} \left(\ux^\top \ubeta^{(d)} \right) = \sum_{l=0}^L \, \xi_{l}^{(d)} \ \psi_l \left(\ux^\top \ubeta^{(d)} \right)$.

\noindent \textbf{Step 2.} Draw $B$ independent samples from joint distribution of $\uR$, \ie 
$$\uR_1, \ldots, \uR_B \sim \text{Dir }(\ualpha^{(d)})$$

\noindent \textbf{Step 3.} Draw $B$ independent samples from the marginal distribution of $T$ hierarchically as follows :
For $i^\prime = 1, 2, \ldots, B$,
\begin{enumerate}
    \item [(i)] Draw $V_{i^\prime} \sim \text{Cat} \left(\pi_1^{(d)}, \ldots, \pi_H^{(d)} \right)$.
    \item [(ii)] Given $V_{i^\prime} = h$, draw $Y_{i^\prime} \sim \mathcal{N}\left(\, g^{(d)}(\ux^\top \ubeta^{(d)}) + \phi_h^{(d)}, \ (\sigma_b^2)^{(d)} + (s^2_h)^{(d)} \,\right)$.
    \item [(iii)] Set $T_{i^\prime} = \exp{\{Y_{i^\prime} \}}$.
\end{enumerate}

\noindent \textbf{Step 4.} 
The SOPs for a given $t \geq 0$, can be expressed in terms of $T$ and $\uR$ by substituting equation (\ref{eq:relative_inc}) in equation (\ref{eq:states}) from Section \ref{s:model} of the main document. We define the corresponding indicator functions $\eta_k$ for each state $k \in \mathcal{K}$, 
\begin{equation*}
\eta_k(t\,;\, T, \uR) = 
\begin{cases}
    \mathds{1} \left( t \ <  T \, R^{(1)} \right), & k = 0 \\
    \mathds{1} \left( T \, \sum_{l=1}^k  {R^{(l)}} \leq \ t \ <  T \, \sum_{l=1}^{k+1}  R^{(l)} \right), & k = 1,\ldots, K-1 \\
    \mathds{1} \left(  T \leq t \right), & k = K
\end{cases}    
\end{equation*}
and estimate the probabilities with their Monte Carlo averages,
\begin{equation*}
\widehat{p}^{\,(d)}_k(t) = \frac{1}B \sum_{i^\prime = 1}^B \eta_k(t \,;\,T_{i^\prime}, \uR_{i^\prime}), \quad k \in \mathcal{K}.
\end{equation*}

\noindent \textbf{Step 5.} The TPs defined as a conditional probability can similarly be estimated by the Monte Carlo method as follows,
\begin{equation*}
    \widehat{p}^{\,(d)}_{rs}(u,t) = \frac{\sum_{i^\prime = 1}^B \eta_s(t + u \,;\,T_{i^\prime}, \uR_{i^\prime}) \ \eta_r(u \,;\,T_{i^\prime}, \uR_{i^\prime})}{\sum_{i^\prime = 1}^B \eta_r(u \,;\,T_{i^\prime}, \uR_{i^\prime})}, \quad r \leq s \,;\  r, s \in \mathcal{K}.
\end{equation*}

\medskip

\section{Additional details on the GAAD data analysis}
\label{s:add_data_analysis}

\subsection{Choice of hyperparameters}
\label{s:hyperparameters}

In this section, we elaborate on the choice of hyperparameters which characterize the prior distributions on our model parameters, used in the GAAD data analysis. 

\vspace{0.3cm}

\noindent \textbf{Regression parameters}: We use a zero-mean Gaussian prior on $\widetilde \ubeta$ and set its prior variance $\sigma_\beta^2 = 100$. However, any arbitrary choice for $\sigma_\beta^2$ would suffice, as long as it is positive. Since $\widetilde \ubeta$ is scaled by its norm, the variance of the induced prior on $\beta$ remains unaffected by the specific choice of $\sigma_\beta^2$. Additionally, one can specify a non-zero mean $\mu$ for $\widetilde \ubeta$ to reflect prior knowledge about certain covariates. To accommodate this, we need to apply a simple modification to the elliptical slice sampling algorithm for posterior sampling of $\widetilde \ubeta$ to adjust for its non-zero mean. Details on implementing this adjustment can be found in Section 3.3 of \cite{murray2010elliptical}.

\noindent \textbf{Monotone link function}:  The length-scale parameter $\ell$ of the Matérn covariance kernel (discussed in Section \ref{s:post_link}) controls the rate of decay of the covariance with the inter-location distances, and is usually chosen to ensure very small correlation between two distant points in the covariate space. The parameter $\nu$ controls the smoothness of the monotone function $g$. We know that the reproducing kernel Hilbert space of a Gaussian process equipped with Matérn covariance kernel with smoothness parameter $\nu$ consists of functions belonging to the Hölder class of smoothness $\nu + 0.5$. Following \cite{Ray2020EfficientBS}, we use default choices of $\nu = 0.75$, with $\ell$ chosen such that the correlation at a maximum possible separation between the covariates equals $0.05$. When using a smooth relaxation of the indicator function in the constrained prior of the basis coefficients in equation (\ref{eq:xi_prior}) of the main document, we set $\eta = 100$. To have a non-informative inverse-gamma $\mathcal{IG}(a_\xi, b_\xi)$ prior on $\tau^2$, both the parameters are set to $0.01$.

\noindent \textbf{Covariance matrix of spatial random effects}: 
The mean of an inverse-Wishart distribution with degrees of freedom $c$ and $m/2 \times m/2$ scale matrix $S$ is $S/(c-m/2-1)$. Here, $c$ controls how informative the prior is. We set $c = m/2+2$ and $S = (E_W - \rho \,W)^{-1}$, such that the prior is proper ($c> m/2+1$), and is loosely centered around a CAR model, where the degree of spatial dependence $\rho$ is set at $0.9$.

\noindent \textbf{Density of errors}: The distinct atoms $(\phi_h, s^2_h)$ specifying the DP mixture of Gaussians are endowed with a conjugate normal-inverse gamma $\mathcal{NIG}(\mu_{\epsilon}, \nu_{\epsilon}, a_{\epsilon}, \lambda_{\epsilon})$ prior, i.e., 
\begin{align*}
    s_h^2 \mid a_{\epsilon}, \lambda_{\epsilon} & \sim \mathcal{IG}(a_{\epsilon}, \lambda_{\epsilon})\,, & \phi_h \mid s^2_h,\, \mu_{\epsilon}, \, \nu_{\epsilon} & \sim \mathcal{N}(\mu_{\epsilon}, s^2_h/\nu_{\epsilon}) \,.
\end{align*}
We set $\mu_\epsilon = 0$, $\nu_\epsilon = 1$, and $a_\epsilon$, $b_\epsilon$ are chosen to have the prior mean and variance of $s_h^2$ as $0.1$ and $10$, respectively.
The parameters of the gamma $\mathcal{G}(a_\gamma$, $\lambda_\gamma)$ prior on the concentration parameter $\gamma$ is chosen so that its mean and variance are $1$ and $10$, respectively. We set the truncation level of the mixture components at $H = 10$.

\noindent \textbf{Dirichlet parameters of the relative disease increment times}: Since the information on the state transition times from the observed data is weak, we use a weakly informative gamma $\mathcal{G}(a_\alpha$, $\lambda_\alpha)$ prior on the Dirichlet parameters $\alpha_k$, $k = 0,1,\ldots, K$, that specifies the distribution of the relative disease increment times $R_{ij}$, so that the prior does not unduly influence the posterior distribution. We choose $a_\alpha$ and $\lambda_\alpha$, such that the prior mean and variance are $1$ and $100$, respectively.

\subsection{Selection of knots in the basis expansion of the monotone link function}

\begin{figure}[htp]
    \centering
    \includegraphics[width=10cm]{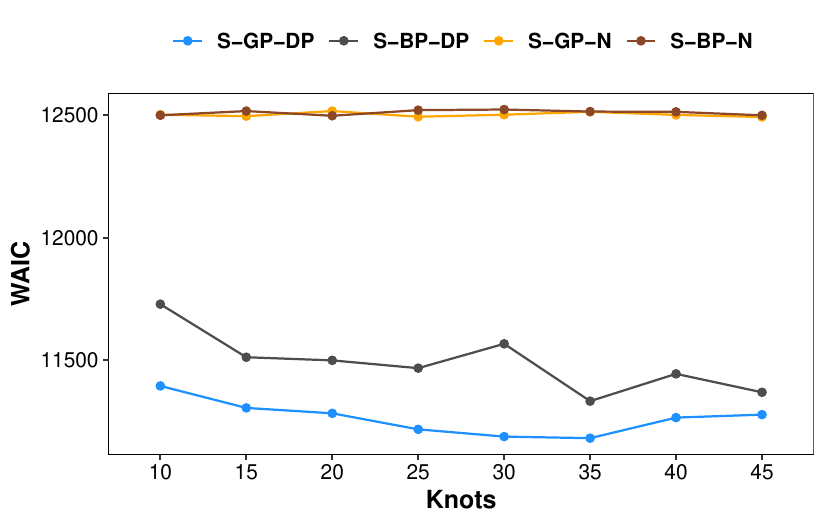}
    \caption{Watanabe-Akaike information criterion (WAIC) for the fitted models plotted against the number of knot points.}
    \label{fig:waic}
\end{figure}

We vary the number of knot points $L \in \{10, 15, 20, \ldots, 45\}$ in the basis expansion of the monotone link function, given in equation (\ref{eq:basis_expansion}) of the main document. By fitting the four competing spatial models to the GAAD data (see Section \ref{s:data_analysis} of the main document for details), we calculate the corresponding values of Watanabe-Akaike information criterion (WAIC) for each knot point. In Figure \ref{fig:waic}, the plotted WAIC values against the number of knots reveal a clear pattern: the S-GP-DP model consistently exhibits the lowest WAIC across all knot points, followed by the S-BP-DP model. Models employing DP errors consistently outperform those with normal errors in terms of WAIC. For the S-GP-DP model, the lowest WAIC occurs at $L=35$, followed by $L=30$ with a difference of less than $10$. Consequently, we select $L=30$ for our analysis of the GAAD data.

\label{s:knots}

\subsection{Plot of model based residuals}
\label{s:residuals}

\begin{figure}[ht]
    \centering
    \includegraphics[width=10cm]{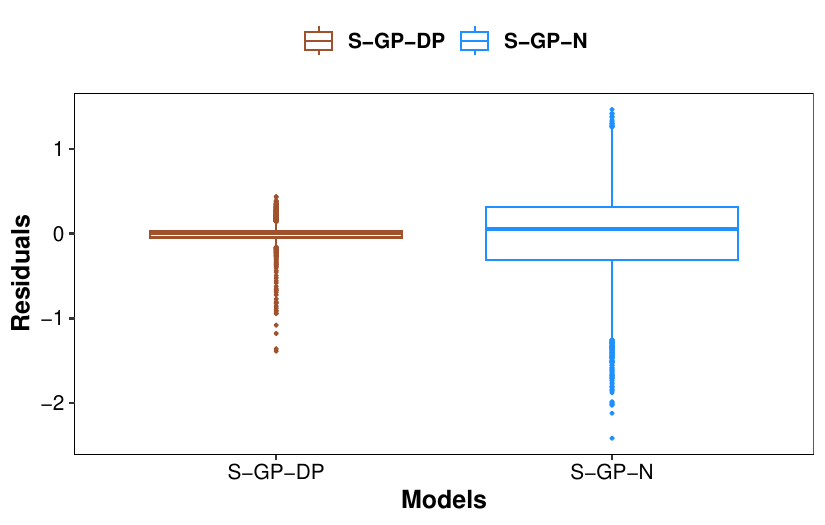}
    \caption{Model based residuals for the spatial models with constrained GP basis for the link function.}
    \label{fig:residuals}
\end{figure}

We define the model based residuals as $e_{ij} = \log T_{ij} - b_{ij} - g(\ux_{ij}^\top \beta) - \phi_{z_{ij}}$ for each $i$, $j$. Figure \ref{fig:residuals} shows boxplots of these residuals for the S-GP-DP and S-GP-N models. It is evident that the model with a normal prior for the error density exhibits greater variability in residuals around zero in comparison to the model employing a DP mixture prior.

\section{Additional simulation results}
\label{s:add_sim}
\begin{figure}
    \centering
    \includegraphics[width=13cm]{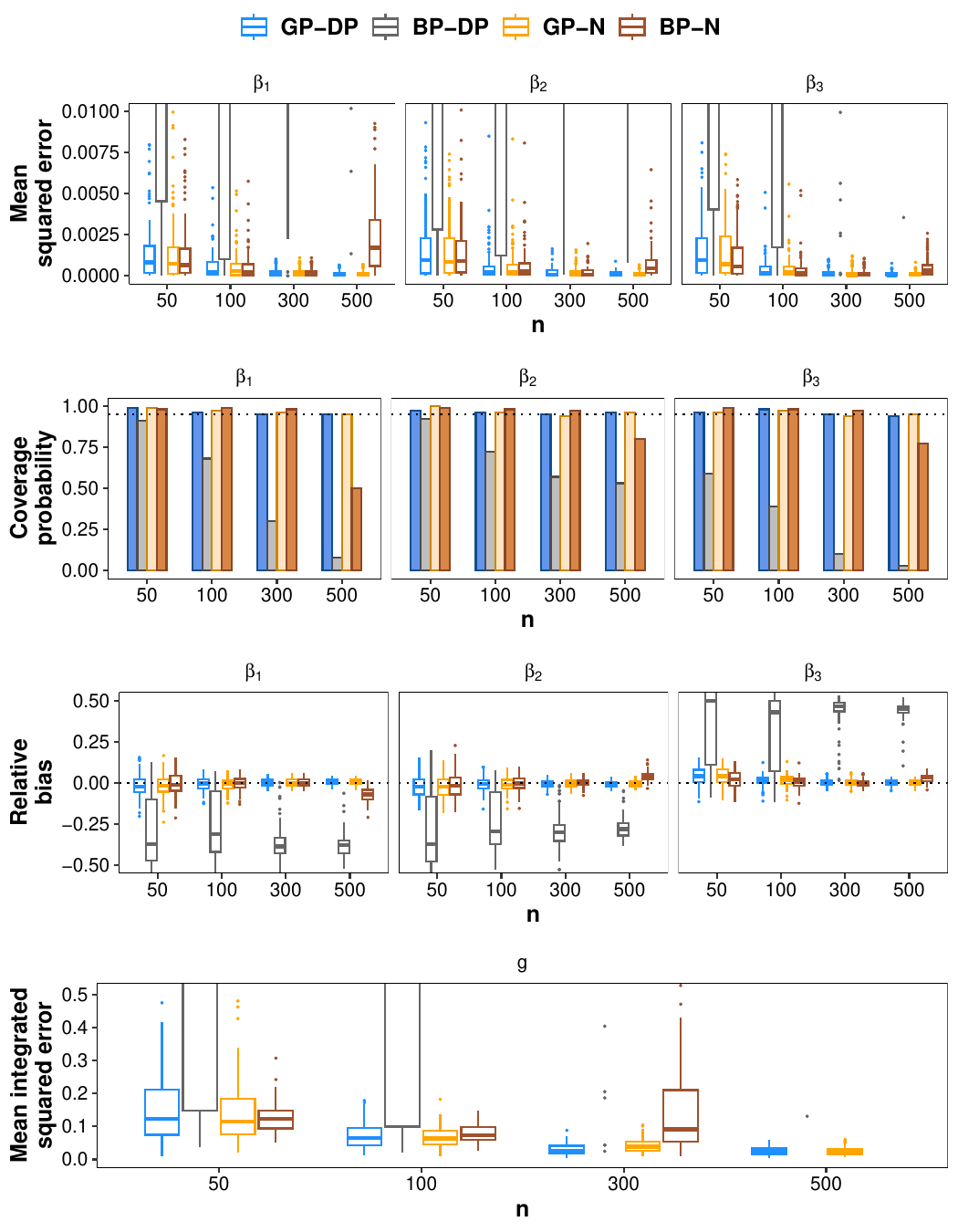}
    \caption{Barplots of coverage probabilities (CP, second), boxplots of mean squared error (MSE, topmost) and relative bias (RB, third) of estimated regression coefficients; and boxplots of mean integrated squared errors (MISE, lowermost) of estimated link function when the true link is $g_2^0$ under Simulation 2: the fitted model is well specified. Plots show variation across 100 simulation replicates.}
    \label{fig:poly1}
\end{figure}

\begin{figure}
    \centering
    \includegraphics[width=13cm]{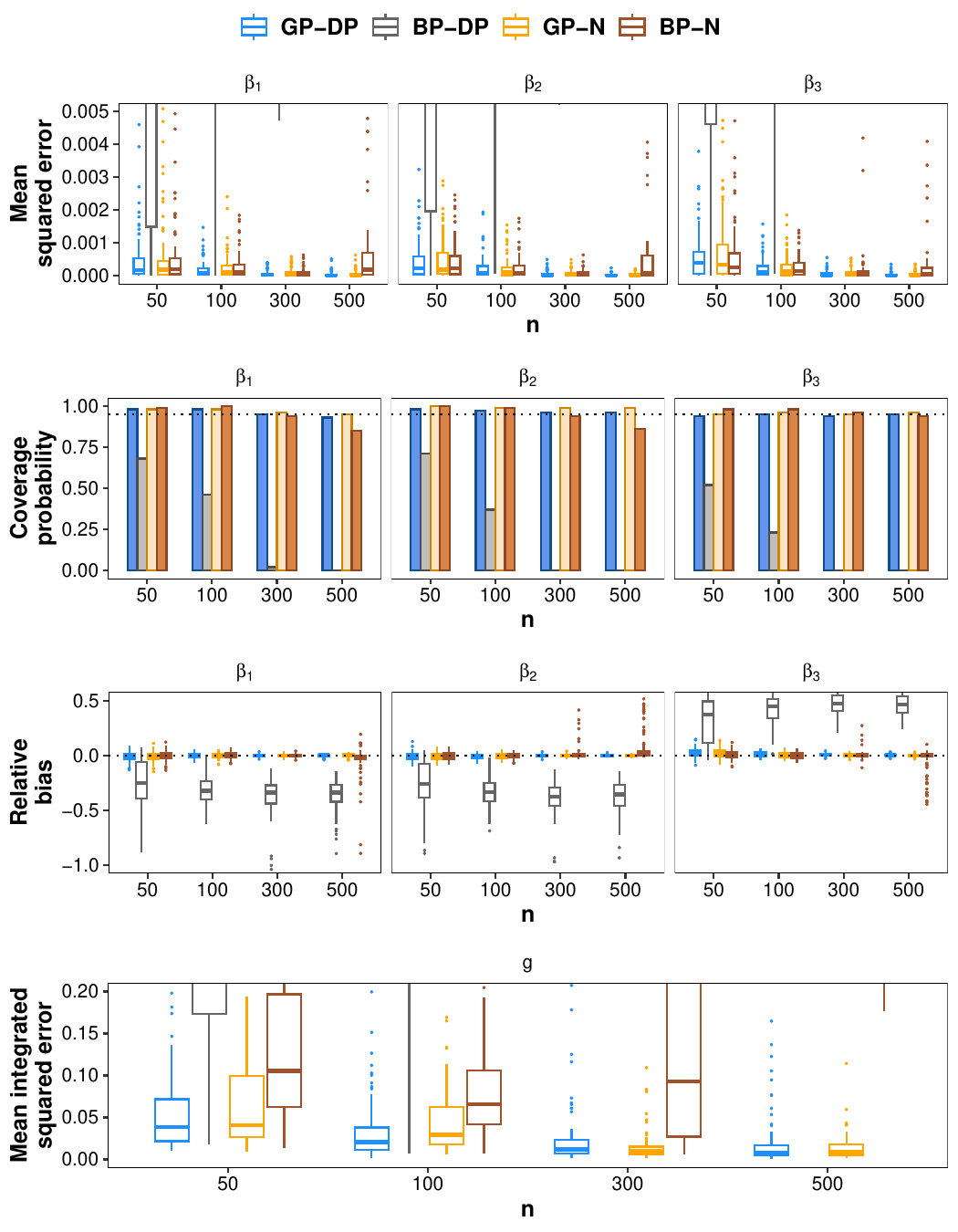}
    \caption{Barplots of coverage probabilities (CP, second), boxplots of mean squared error (MSE, topmost) and relative bias (RB, third) of estimated regression coefficients; and boxplots of mean integrated squared errors (MISE, lowermost) of estimated link function when the true link is $g_2^0$ under Simulation 2: the fitted model is misspecified. Plots show variation across 100 simulation replicates.}
    \label{fig:poly2}
\end{figure}

In this section, we present the results of Simulations 1 and 2 when the underlying true link function is $g_2^0$. Boxplots of the mean squared error (MSE), relative bias (RB) and mean integrated squared error (MISE) and barplots of the coverage probability (CP) across the 100 replicates over increasing $n$ under Simulations 1 and 2 are shown in Figures \ref{fig:poly1} and \ref{fig:poly2}, respectively.

Under Simulation 1, where the model is correctly specified, both the S-GP-DP and S-GP-N models demonstrate superior estimation performance compared to models employing the BP basis. The boxplots of MSE, RB, and MISE exhibit a decreasing trend as sample sizes increases, indicating that posterior estimates from models with constrained GP basis are consistent. The desired $95\%$ coverage is attained by each $\hat \beta_j$ for both small and large sample sizes. The S-BP-N model shows comparable performance for sample sizes (50, 100, 300) but the performance deteriorates for sample size 500. The S-BP-DP model demonstrates uniformly poor performance for all sample sizes.

Under Simulation 2, where the model is misspecified, the S-GP-DP and S-GP-N models consistently outperform the S-BP-DP and S-BP-N models. The GP models deliver satisfactory estimation performances in terms of MSE, RB and MISE. The GP models attain the desired $95\%$ coverage across all sample sizes. While S-BP-N maintains the nominal coverage rate, the coverage of S-BP-DP shows a decline as the sample size increases, dropping to $1\%$ for $n = 500$. The regression parameter estimates from S-BP-N remain robust for all sample sizes, but the estimation of the link function deteriorates with an increase in sample size.  Estimation of the link function and the regression parameters are seen to be poor under the S-BP-DP model. In summary, estimates from the GP models are seen to be robust under model misspecification. Estimation of the link function under the S-BP-N model tends to suffer for large sample sizes, while the S-BP-DP model delivers poor performances across all sample sizes.

\end{document}